\documentclass[journal=jpccck, manuscript=article, layout=traditionffal]{achemso}
\SectionNumbersOn

\usepackage[outlinesitem]{mypaper}
\proofreadfalse

\title[PlasmPropMonoBi]{Plasmonic Properties of Close-packed Metallic Nanoparticle Mono- and Bilayers}

\author{Bruno G. M. Vieira}
\affiliation{Universidade Federal do Ceará, Department of Physics, Fortaleza, Ceará, 60455-760 Brazil.}
\affiliation{Freie Universität Berlin, Department of Physics, Arnimallee 14, D-14195 Berlin, Germany}
\altaffiliation{Contributed equally to this work}
\email{bruno@fisica.ufc.br}

\author{Niclas S. Mueller}
\altaffiliation{Contributed equally to this work}
\affiliation{Freie Universität Berlin, Department of Physics, Arnimallee 14, D-14195 Berlin, Germany}

\author{Eduardo B. Barros}
\affiliation{Universidade Federal do Ceará, Department of Physics, Fortaleza, Ceará, 60455-760 Brazil.}

\author{Stephanie Reich}
\affiliation{Freie Universität Berlin, Department of Physics, Arnimallee 14, D-14195 Berlin, Germany}


\begin{document}

This document is the unedited Author's version of a Submitted Work that was subsequently accepted for publication in The Journal of Physical Chemistry C, copyright \textcopyright American Chemical Society after peer review. To access the final edited and published work see the following link: \href{https://pubs.acs.org/doi/10.1021/acs.jpcc.9b03859}{https://pubs.acs.org/doi/10.1021/acs.jpcc.9b03859}.

\newpage

\begin{abstract}
	The self-assembly of metallic nanoparticles is a promising route to metasurfaces with unique
	properties for many optical applications, such as surface-enhanced spectroscopy, light manipulation, and sensing. We present an in-depth theoretical study of the optical properties of mono- and bilayers assembled from gold and silver nanoparticles. With finite-difference time-domain simulations, we predict the occurence of two plasmon modes, a bright and a dark mode, which exhibit symmetric and antisymmetric dipole configurations between the layers, respectively. 
	 The dark mode resonance energy depends sensitively on the size of the particles and the interparticle gaps. Hotspots with a nearfield intensity enhancement of up to 3000 are expected, which, together with the fact that the dark mode is roughly four times narrower than the bright mode, reveals how promising these materials are for spectroscopy purposes.  
\end{abstract}

\section{Introduction \label{sec:intro}}

	Noble-metal nanoparticles (NMNPs) have been studied for many years due to the unique way in which they interact with light. The interaction is mediated by the free electrons inside the nanoparticles (NPs) that are collectively excited by the incident light, which is known as a localized surface plasmon resonance (LSPR)\cite{EtchegoinBook2009,MaierBook2007}.
	These resonances are strongly dependent on the structural parameters of the nanoparticles, such as shape and size\cite{Kelly2003}. 
	Due to the strong electric nearfield generated by those excitations in the vicinity of the NMNPs, there is a remarkable enhancement of the optical properties of any material that is close to the metal surface. Consequently, the NMNPs are considered to be excellent systems for applications in high-resolution imaging beyond the diffraction limit\cite{Gramotnev2010} and in spectroscopy techniques, such as, fluorescence\cite{Flauraud2017}, infrared absorption\cite{Hartstein1980} and surface-enhanced Raman scattering (SERS)\cite{Benz2016,Zhang2013,Kneipp1997,Moskovits1985}. In fact, there are many other areas in which the NMNPs may be employed, for instance, in optoelectronic devices, as they can shape the light patterns on the nanoscale\cite{Schuller2010}, in biological and chemical sensing\cite{Mayer2011}, in signal-processing systems\cite{Gramotnev2010} and in photovoltaic devices\cite{Atwater2010}.
	
	Beyond single nanoparticle systems, it is also possible to cluster nanoparticles together to form more complex structures, bringing up a whole branch of possibilities of novel materials. Due to the coupling between the plasmon modes of each individual particle, these ensembles of NMNPs are known to have optical properties that differ from those of individual nanoparticles and their corresponding bulk materials\cite{Ross2016,Nie2009}. Their optical properties can be additionally tuned by changing the particles distance and arrangement. The self-assembly of NMNPs is an attractive synthesis technique for close-packed ensembles of nanoparticles. The control over structural parameters has been greatly improved in the past years\cite{Julin2018,Schulz2017,Gong2017,Hanske2017,Klinkova2014,Grzelczak2010,Tao2008}. For instance, Schulz et al. \cite{Schulz2017} were able to synthesize large films (hundreds of $\mu$m$^2$) of close-packed gold nanospheres with particle sizes of up to 46~nm and interparticle distances of 2~nm with a precision of 0.5~nm. Besides being a relatively low cost synthesis method in comparison to electron-beam lithography, it is very versatile, as many different molecules can be used as ligands to achieve various gap sizes and join together particles with a wide range of sizes. In addition, as those nanoparticle films extend over length scales far beyond the diffraction limit, optical measurements can be easily done on them without the need of complex experimental apparatus\cite{Mueller2018,Mueller2018a}. 
	
	\condcolor{red}{ 
	The optical properties of close-packed ensembles of metallic nanoparticles has been subject in a number of theoretical works. Tao et al.\cite{Tao2008} considered polyhedral silver nanoparticles colloids of up to three layers and investigated their optical response in both low density and high density configurations. 
	Alaeian et al.\cite{Alaeian2012} studied spherical and octahedral nanoparticles superlattices and showed that they behave as metamaterials with magnetic response.
	Our recent studies also provided some theoretical analysis on the self-assembly of NMNPs, but the focus was mostly on experimental measurements and potential applications of these materials\cite{Mueller2018,Mueller2018a}. 
	Furthermore, there is a number of theoretical studies on arrays of metal-insulator-metal particles (MIMs)\cite{Chang2012,Wang2017,Frederiksen2013,Verre2015,Cai2016}. Although the MIMs are synthesized in a process different from the self-assembly, their optical response has similarities with the response of close-packed nanoparticle bilayers, which makes them great to compare with.   
	Although the interest in NMNP layers increases and the development of self-assembly techniques improves at a rapid pace, there is still a lack of a systematic theoretical investigations on the role of geometrical parameters on the plasmonic properties.}

	In this paper, we provide a thorough theoretical framework about the plasmonic properties of layered structures of gold and silver nanoparticles. We will focus on the properties of mono- and bilayers, which are the fundamental building blocks for more complex layered structures. 
	\condcolor{red}{ 
	We predict that the monolayers have one plasmon mode, characterized by a non-zero net dipole moment (bright mode), and the bilayers enable the excitation of a bright mode and a second mode that has a vanishing dipole moment (dark mode). This dark mode is activated due to field retardation, that induces an anti-parallel dipole configuration between the layers. 
	The plasmonic modes can be excited by illumination with a plane electromagnetic wave at normal incidence and the optical response is polarization insensitive.
	We provide a systematic study on the influence of the structural parameters, including particles size (diameter), their separation (gap) and particle arrangement. From the size analysis, we were able to predict the minimum diameter for which the dark mode is activated. By an analysis of the local nearfield enhancement, we demonstrate the potential of the nanoparticle layers for surface-enhanced spectroscopy. 
	}
	
	This paper is organized as follows. We first introduce the details of the theoretical approach used throughout the paper (Sect. \ref{sec:methods}). In Sect. \ref{sec:monolayer}, we show the results for the monolayers with emphasis on the effect of particle separation and size. The simulations for the bilayers are presented in Sect. \ref{sec:bilayer}. We study the dependence of the plasmon modes on structural parameters, such as particles size, particle separation and layer stacking. Finally, we study the nearfield enhancement of the bilayers and offer conclusions (Sect. \ref{sec:conc}).

\section{Methods \label{sec:methods}}
	
	\begin{figure}[!ht]
		\centering
		\includegraphics[width=.4\linewidth]{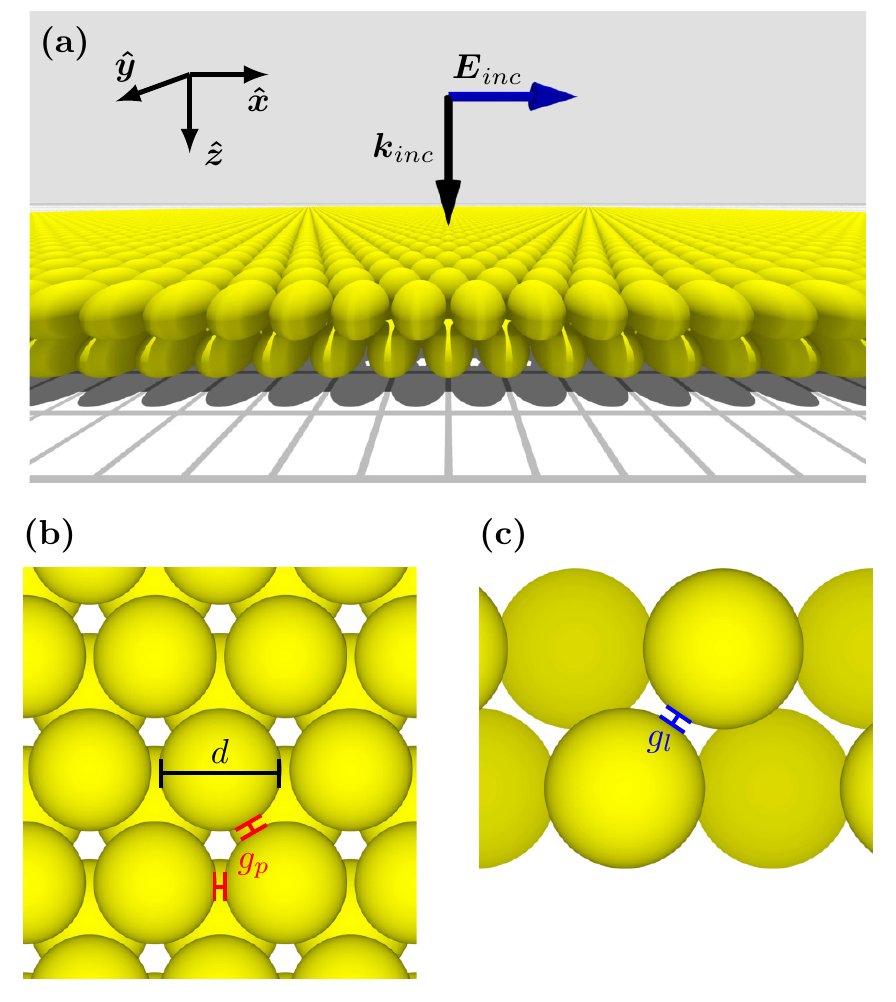}
		\caption{Example of a bilayer of gold nanospheres investigated in this study in three different views: {\bfseries (a)} perspective view, {\bfseries (b)} top view (Plane $xy$) and {\bfseries (c)} side view (Plane $yz$). $\vect{k}_{inc}$ and $\vect{E}_{inc}$ indicate the propagation and polarization directions of the incident light, respectively. In {\bfseries (b)} and {\bfseries (c)}, the particles diameter ($d$), the gap between neighboring particles in the layers ($g_p$) and the gap between the layers ($g_l$) is indicated.}
		\label{fig:System_PerspecView}
	\end{figure}

	\condcolor{red}{ 
	In this work, we consider mono- and bilayers of close-packed metallic nanospheres of equal size, as illustrated in Fig. \ref{fig:System_PerspecView}. In all simulations, we considered vacuum as the surrounding medium ($\epsilon_m=1$) and a linearly polarized plane wave with propagation normal to the layers as the light source. The polarization direction was set as the $x$ direction, although any other polarization direction would lead to the same spectra because of the hexagonal symmetry of the nanoparticle lattice.
	}
	\condcolor{blue}{ 
	Throughout the paper, we will use the following notation regarding the layers structural parameters: $d$ is the particles diameter, $g_p$ is the the edge-to-edge spacing (gap) between particles within one layer and $g_l$ is the gap between the layers [see Figs. \ref{fig:System_PerspecView}(b) and (c)]. The variable $g$ without subscript is used when $g_p$ and $g_l$ have equal values and are simultaneously changed ($g\equiv g_p=g_l$). Regarding the spectral features of the plasmon modes, we will denote their excitation energy (peak position) as $E$ and their full width at half maximum (FWHM) as $\Gamma$.
	}

	All simulations were performed with Lumerical FDTD Solutions, a commercial software package. The monolayers were considered as two-dimensional crystal lattices and simulated by defining a square unit cell and using periodic boundary conditions. The bilayers were implemented by adding more NPs to the unit cell such that the NPs arrangement determines the layers stacking (align and hcp). Mesh-override regions of 0.25~nm were used for systems with particle diameters below 15~nm and gap sizes below 3~nm while, for larger diameters and gaps, an override mesh size of 1~nm was used. Two power monitors were included; one behind the layers to capture the transmittance $T$, and another one behind the source for the calculation of the reflectance $R$. The absorbance $A$ was later computed as $A=1-T-R$. A plane-wave source was used for the layers simulations while a total-field scattered-field source was used to calculate the cross-section (CS) of nanoparticle dimers. The gold and silver dielectric functions were implemented by fitting the experimental data provided by Johnson and Christy\cite{Johnson1972} and by Palik\cite{PalikBook1985}, respectively. Convergence tests were performed by varying the mesh sizes in order to assure the accuracy of the simulation results. The polarization density $\vect{P}$ was calculated from the simulation data for the electrical current density $\vect{J}$ by using the relation $\vect{J}=\frac{d\vect{P}}{dt}$ and assuming a harmonic time dependence. The spatial average over the unit cell region of each layer ($<\vect{P}>$) was then computed in order to observe just the mean behavior of the polarization.
	
\section{Nanoparticle Monolayer \label{sec:monolayer}}

	\begin{figure}[h]
		\centering
		\begin{tikzpicture}
			\node (plot) {\includegraphics[scale=.8,trim=0 0 70 0,clip]{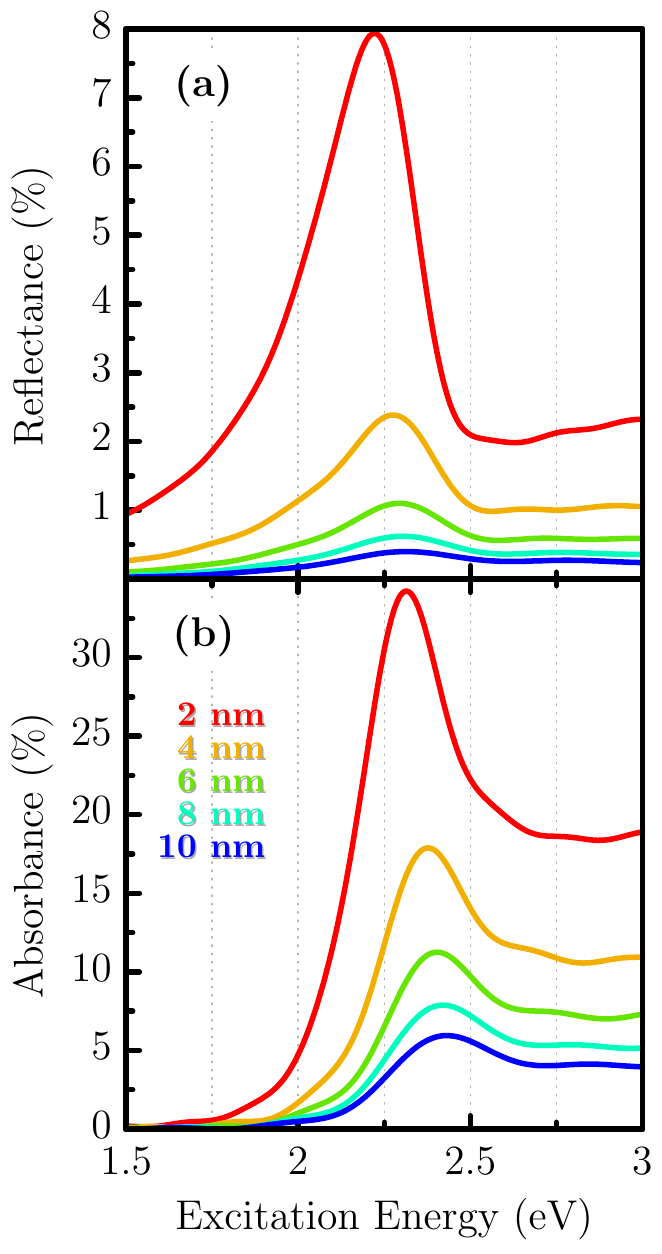}};
			\node[inner sep=0,below left] (illust) at ([shift={(-.4,-.5)}]plot.north east)
			{\includegraphics[width=.079\paperwidth,trim=750 0 750 0,clip]{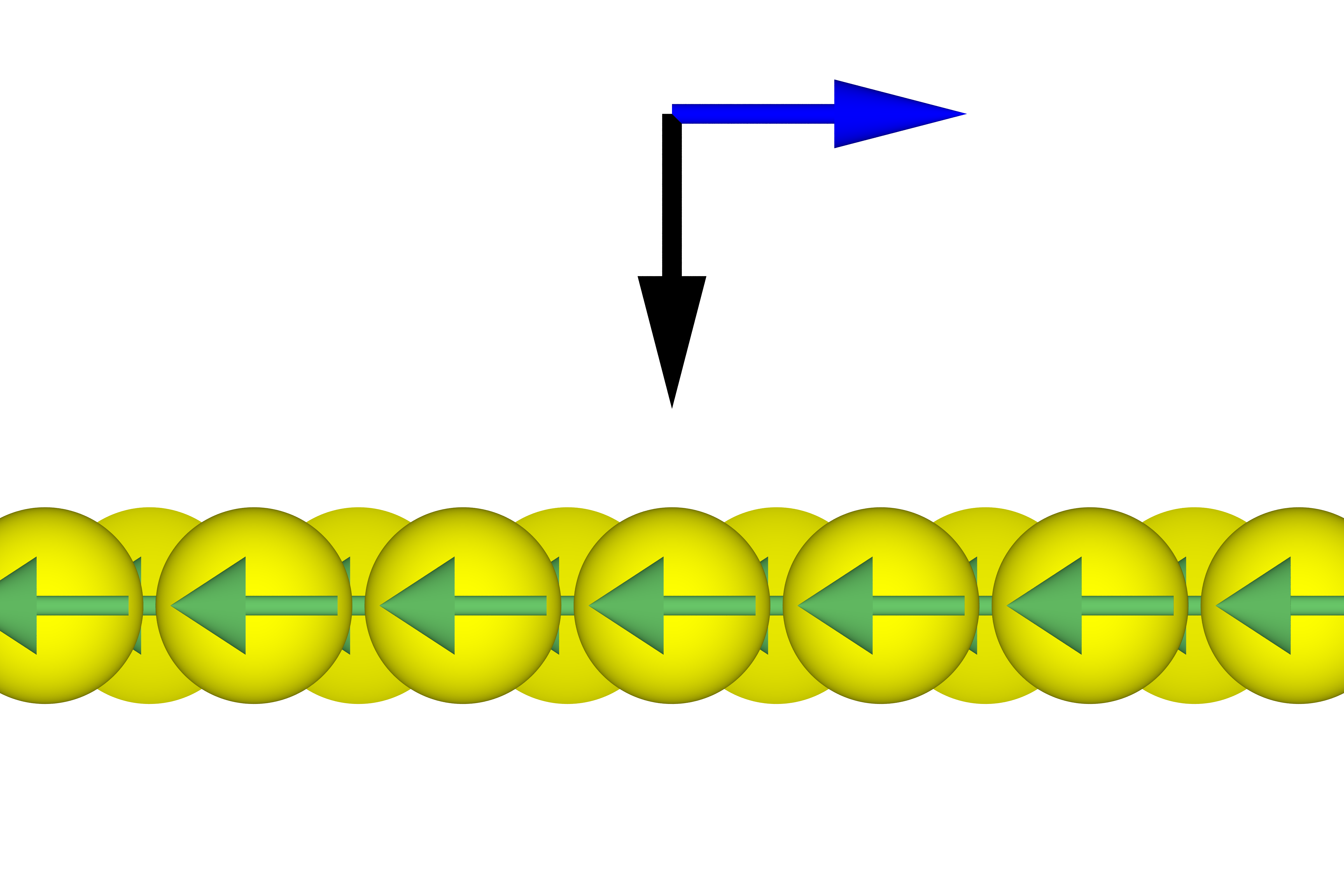}};
			\foreach \pos/\text in {{(.45,.35)}/E,{(-.31,0.35)}/k} \node[fill=white,scale=.6,inner sep=1pt] at ([shift={\pos}]illust.center){$\vect{\text}_{inc}$};
		\end{tikzpicture}
		\caption{{\bfseries (a)} Reflectance and {\bfseries (b)} absorbance spectra calculated for monolayers of gold nanospheres with diameter $d=10$~nm. The particles are arranged in a hexagonal lattice and the spectra for different gap sizes ($g$) are shown.}
		\label{fig:Au_D10NL1_SwGap}
	\end{figure}
	
	We start our investigation by analyzing the simplest case, i.e., a monolayer of hexagonally packed metallic nanospheres. Figure \ref{fig:Au_D10NL1_SwGap} shows the absorbance and reflectance spectra of a single layer of gold nanospheres with $d=10$~nm. 
	\condcolor{red}{ 
	Its optical response is polarization insensitive as the monolayers have a hexagonal symmetry (Point Group $D_{6h}$) \cite{Jorio2017}.
	}
	The distinct curves in each panel indicate the influence of the gap between the particles ($g$). A plasmon mode is excited around 2.35~eV in all the cases and increases in intensity as $g$ decreases. 
	This mode corresponds to the dipole-active mode in which all nanoparticles have parallel dipole moments of similar magnitude (see inset of Fig. \ref{fig:Au_D10NL1_SwGap}).
	The absorbance spectrum is dominant over reflectance, because small spheres are poor scatterers of light as they do not have a large amount of material to prevent light from being transmitted. There is a slight redshift that stems from the increased coupling between the particles as $g$ is reduced. 
	\condcolor{red}{ 
	Similar shifts has been observed previously in the optical response of nanosphere colloids\cite{Liz-Marzan2006} and metallic nanoparticle oligomers, such as spherical and elliptical dimers\cite{Ross2016,Halas2011,Lassiter2008,LeRuInBook2009,Ghosh2007,Romero2006,Zhong2004,Su2003,Jensen1999}, nanorod dimers and bowtie antennas\cite{Peyskens2015}.
	}
	

	
	\begin{figure}[!ht]
		\centering
		\begin{tikzpicture}
			\def\h{1.1cm}
			\def\y{-1.5cm}
			\node (plot) {\includegraphics[page=1]{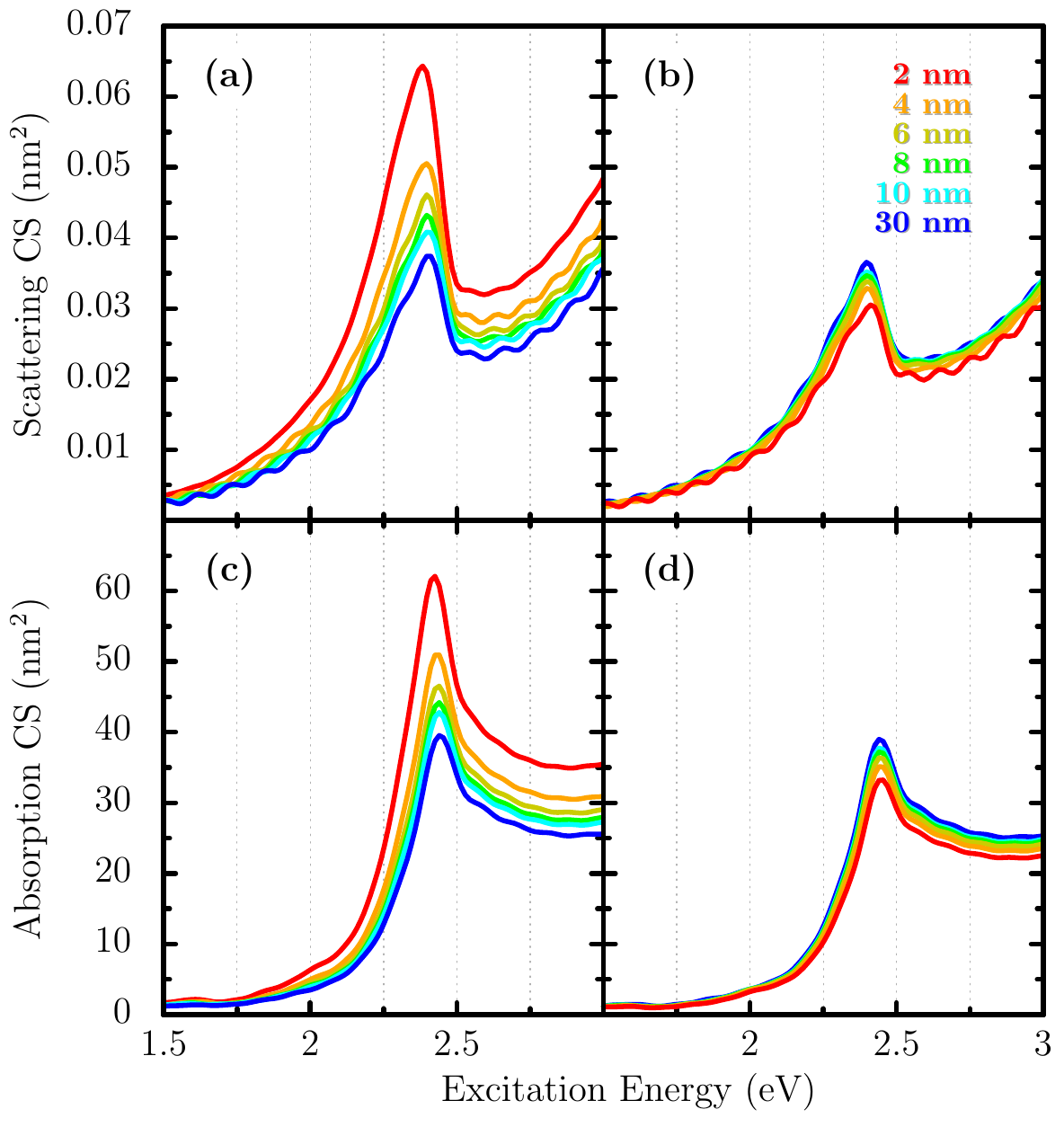}};
			\foreach \p/\pos in {2/{(-3,4.05)},3/{(2.15,4.05)}}\node[inner sep=0,line width=1mm] (insets) at ([shift={\pos}]plot.center) {\includegraphics[page=\p,scale=.38,trim=4 14 170 198,clip]{Figs/Oligo_AuJCSphere_D10NP2_BG1_TFSFLinearlight_SwGap-Paper}};
			\foreach \pos/\text in {{(-2.75,\y)}/B3u,{(2.3,\y)}/B2u} \node at ([shift={\pos}]plot){\includegraphics[height=\h,trim=0 400 0 400,clip]{Figs/pov-ray/Dimer_Mode\text}};
		\end{tikzpicture}
		\caption{Absorption and scattering cross-sections (CS) of a dimer of nanospheres with $d=10$~nm as function of the gap size between the nanoparticles. The left panels show the {\bfseries (a)} scattering and {\bfseries (c)} absorption CS for the light polarization along the dimer axis while the right panels show the {\bfseries (b)} scattering and {\bfseries (d)} absorption CS for the polarization perpendicular to it. The insets in {\bfseries (a)} and {\bfseries (b)} show the zoomed spectral range around the resonance peaks and, in {\bfseries (c)} and {\bfseries (d)}, the insets indicate the dipole configuration of each particle for the different polarizations.}
		\label{fig:Dimer_D10_SwGap}
	\end{figure}

	To get a deeper understanding of the spectra of the nanoparticle monolayers, we draw a comparison to the simple case of a dimer of spheres illuminated from the top. Figure \ref{fig:Dimer_D10_SwGap} exhibits the absorption and scattering cross-sections (CS) of a dimer of particles with $d=10$~nm (same $d$ as in Fig. \ref{fig:Au_D10NL1_SwGap}). Simulations for light polarization along and perpendicular to the dimer axis are presented in order to account for the different angles between the incident light polarization and each of the particle pairs in the layer.
	\condcolor{red}{ 
	As expected, there is a redshift with decreasing $g$ for light polarization along the dimer axis, Fig. \ref{fig:Dimer_D10_SwGap}(a), and a blueshift for perpendicular polarization, Fig. \ref{fig:Dimer_D10_SwGap}(b). The shifts are overall small due to the small size of the nanoparticles\cite{Ross2016,Halas2011,LeRuInBook2009,Ghosh2007,Jensen1999}.}
	For better visualization, the zoomed spectral range of the scattering CS spectra are shown as insets in Figs. \ref{fig:Dimer_D10_SwGap}(a) and (b).
	The parallel polarization causes a binding plasmon mode in the structure, i.e., the induced dipole configuration makes the particles to have an attractive interaction, see inset in Fig. \ref{fig:Dimer_D10_SwGap}(c). On the other hand, the perpendicular polarization excites an anti-binding mode (repulsive interaction between the particles), as in the inset of Fig. \ref{fig:Dimer_D10_SwGap}(d). As $g$ is reduced, the binding (anti-binding) coupling increases, which causes the mode to shift towards lower (higher) energies, i.e., a redshift (blueshift).
	\condcolor{red}{ 
	The dipoles of binding mode are oriented parallel to the dimer axis while the dipoles of the anti-binding mode are perpendicular to it. This orientation difference indicates that the binding coupling is more intense than the anti-binding coupling, an effect that is similar to the observed difference in coupling strength for $\sigma$ and $\pi$ bonds in molecules. It is expected then that the binding mode is subjected to a redshift that is larger in magnitude than the blueshift of the anti-binding mode.}
	
	Back to the monolayer of nanoparticles, there are pairs of particles with a dipole configuration similar to each of the dimer modes shown in Fig. \ref{fig:Dimer_D10_SwGap} (binding and anti-binding modes) while others have a dipole configuration that is a combination of these two modes depending on their orientation (a more detailed description of this coupling will be given further on). Therefore, the fact that the monolayer absorbance shows a redshift with decreasing $g$ similarly to exciting the dimer with polarization along its axis means that the overall binding contribution of the particles affects the energy stronger than the total anti-binding contribution. The behavior was also observed for silver nanoparticle monolayers (see Fig. S1 in the Supporting Information).
	
	\begin{figure}[h]
		\centering
		\includegraphics[scale=.8,trim=0 0 70 0,clip]{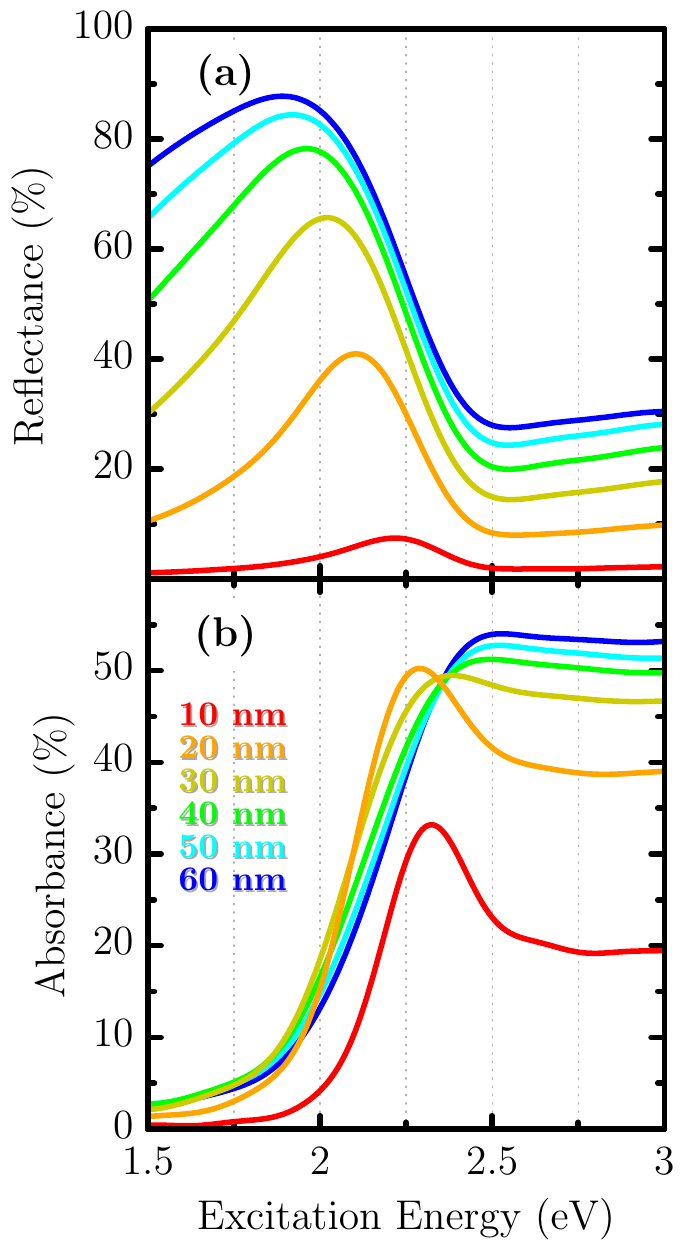}
		\caption{Diameter dependence of the {\bfseries (a)} reflectance and {\bfseries (b)} absorbance for gold nanosphere monolayers with $g=2$~nm.}
		\label{fig:G2NL1_SwDiam}
	\end{figure}
	
	Let us now evaluate how the particles diameter $d$ affects the optical properties. In Fig. \ref{fig:G2NL1_SwDiam} the absorbance and reflectance spectra are displayed for various $d$ using a gap size $g=2$~nm. The plasmon mode observed in Fig. \ref{fig:Au_D10NL1_SwGap} is also excited in larger spheres. As $d$ increases, an absorbance tail above 2.25~eV rises and overlaps the plasmonic peak. This behavior stems from the increase of the interband transitions of gold, caused by the increasing amount of material, in the case of the larger spheres.  
	As LSPRs are surface effects and the interband transitions are related to the NP volume\cite{Pinchuk2004,MaierBook2007}, the damping of the plasmon mode is indeed expected due to the decrease of the NPs surface-to-volume ratio with increasing $d$. 
	 The volume increase is also responsible for the overall enhancement in the reflectance spectra that goes from less than 10\% for $d=10$~nm to a peak of almost 90\% for $d=60$~nm, see Fig. \ref{fig:G2NL1_SwDiam}(a). For the absorbance in Fig. \ref{fig:G2NL1_SwDiam}(b) we note that the spectrum at $d=20$~nm is slightly out of the trend with respect to the larger diameters, indicating that the plasmon mode is still not completely damped away by the interband transitions for 20~nm diameter particles. 
	

\section{Nanoparticle Bilayer \label{sec:bilayer}}

	\begin{figure}[!ht]
		\centering
		\begin{tikzpicture}
			\node (plot) {\includegraphics[page=1,scale=.75]{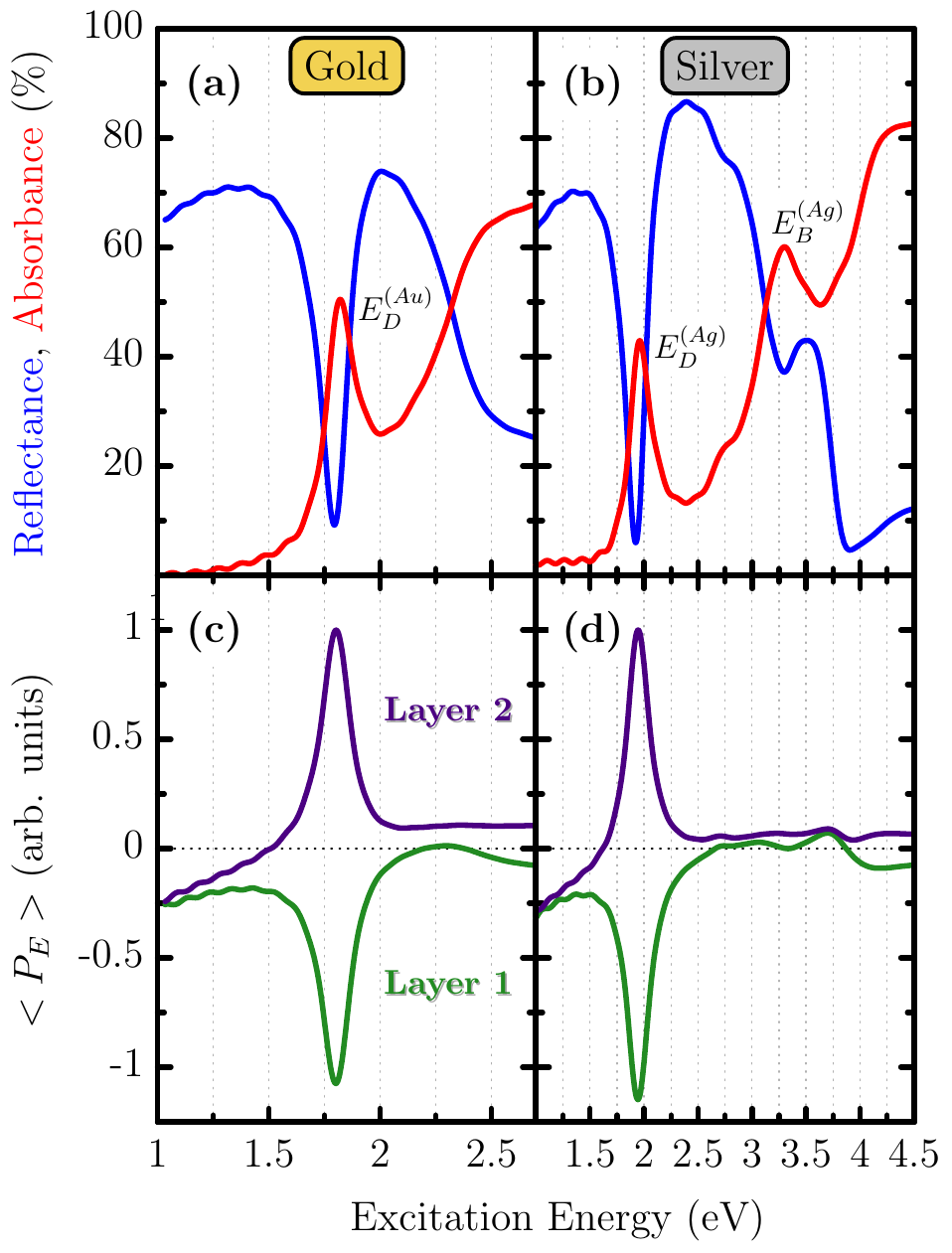}};
			\node[inner sep=0,below left] (illust) at ([shift={(-.68,.2)}]plot.east)
			{\includegraphics[width=.07\paperwidth,trim=750 0 750 0,clip]{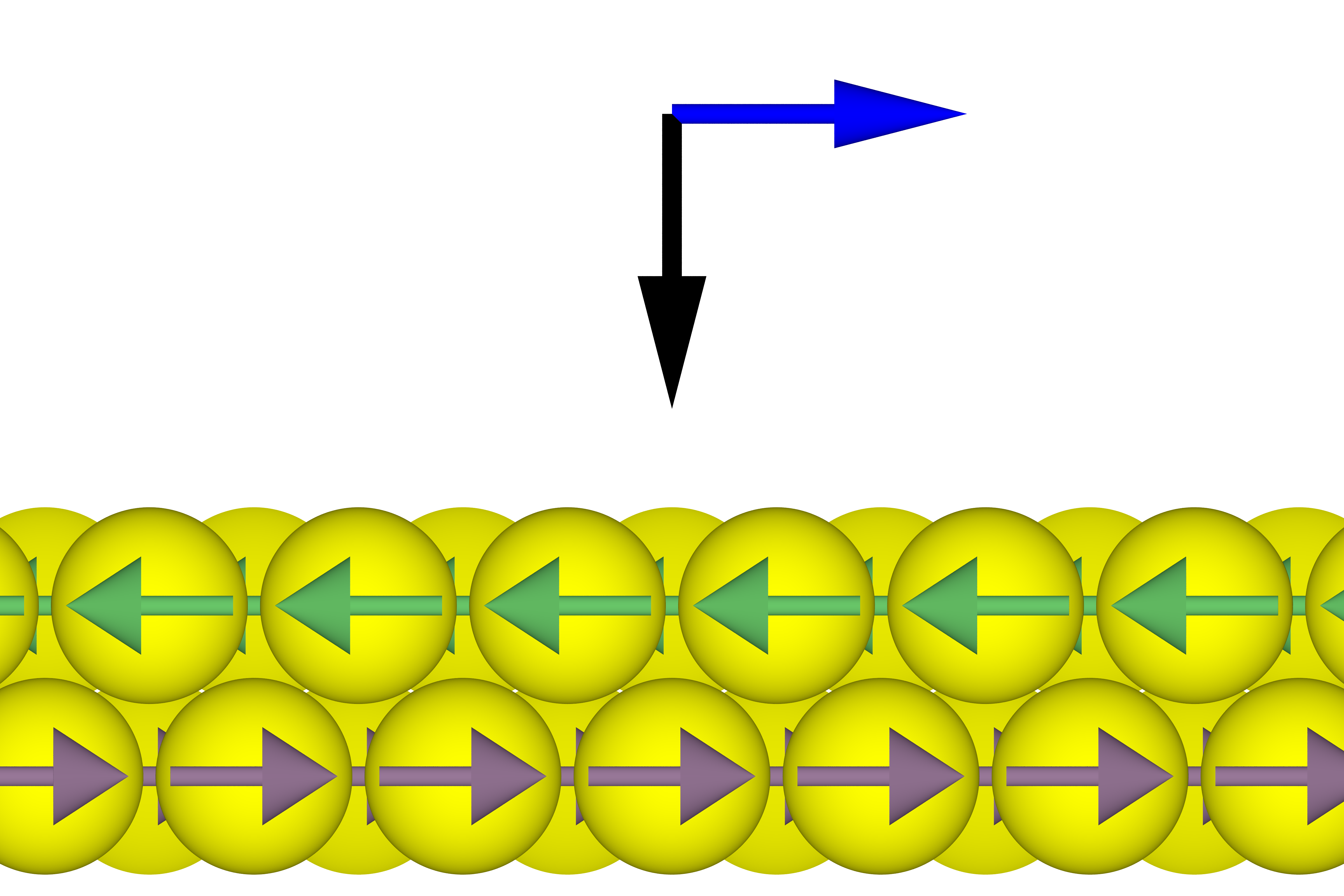}};
			\foreach \pos/\text in {{(.45,.30)}/E,{(-.31,0.35)}/k} \node[fill=white,scale=.6,inner sep=1pt] at ([shift={\pos}]illust.center){$\vect{\text}_{inc}$};
		\end{tikzpicture}
		\caption{Absorbance (red) and reflectance (blue) spectra of a bilayer system of {\bfseries (a)} gold and {\bfseries (b)} silver NPs ($d=50$~nm, $g=2$~nm, hcp stacking).  The mean polarization density $<P_E>$ in each layer for {\bfseries (c)} gold and {\bfseries (d)} silver as function of the excitation energy is also shown. The subscript $E$ means we are considering the component of $<\vect{P}>$ that is parallel to the incident electric field $\vect{E}_{inc}$ (all other components of $<\vect{P}>$ are null). The inset in {\bfseries (d)} shows the dipole configuration of the bilayer when the mode $D$ is excited.
		}
		\label{fig:Bilayer_Default-ARPol}
	\end{figure}
	
	Figure \ref{fig:Bilayer_Default-ARPol}(a) shows the absorbance and reflectance for a gold bilayer of nanospheres with $d=50$~nm and $g=2$~nm.
	A prominent absorbance peak emerges at $E^{(Au)}_D=1.82$~eV together with a reflectance dip at 1.79~eV. The peak has a FWHM of $\Gamma^{(Au)}_D=146$~meV. The dipole-active plasmonic mode ($B$) observed for the monolayer structures is no longer visible in Fig. \ref{fig:Bilayer_Default-ARPol}(a). 
	In the bilayers, the interband transitions of gold are further increased in comparison to the monolayers due to their larger amount of gold, similarly to the effect observed for the monolayers with larger NPs. Therefore, the dipole-active mode is completely damped.
	A very similar optical response was found for bilayers of silver nanoparticles with an absorbance peak at $E^{(Ag)}_D=1.97$~eV and a reflectance dip at 1.93~eV.
	The plasmon mode $B$ is observed as an absorbance peak and a reflectance dip around $E^{(Ag)}_B=3.30$~eV. 
	This peak is expected because the interband transitions of silver occur at energies larger than $3.8$~eV\cite{Pinchuk2004}, which means that less damping should occur to this plasmonic mode in silver in comparison to gold.
	Regarding the modes linewidth, the peak at $E^{(Ag)}_B$ has a FWHM of $\Gamma^{(Ag)}_B=830$~meV, whereas the peak at $E^{(Ag)}_D$ has $\Gamma^{(Ag)}_D=217$~meV,  i.e., it is almost four times narrower (see Fig. S3 in the Supporting Information). 
	
	
	To identify the plasmon modes that are being excited, we investigate the charge distribution in the nanoparticles of each layer. We calculate the mean polarization density $<P_E>$ of each layer as a function of the excitation energy (Figs. \ref{fig:Bilayer_Default-ARPol}c and d). A peak in both layers is observed at $E^{(Au)}_D$, but the sign of $<P_E>$ is opposite in the two layers. This indicates that the associated absorbance peak and reflectance dip are due to the excitation of a plasmon mode with anti-parallel dipole moments between the two layers. 
	The net dipole moment of the bilayer vanishes, which means that this is a dark plasmon mode\cite{Mueller2018,Chu2009}. 
	Dark plasmons have become more relevant in the last years since they exhibit little or no radiative damping and, consequently, longer lifetimes and narrower resonances compared to bright plasmons\cite{Zhang2016,Yanai2014,Sancho-Parramon2012,Chang2012}.
	This expectation agrees nicely with the remarkable difference between $\Gamma^{(Ag)}_D$ and $\Gamma^{(Ag)}_B$ in the bilayers.
	
	Dark plasmons normally do not interact with far-field radiation. However, various approaches have been proposed (and partly demonstrated) for their excitation, such as breaking the symmetry of a system to turn the dark plasmon slightly bright\cite{Humphrey2016,Panaro2014,Chuntonov2011,Chu2009}, using light sources with spatial polarization profiles  \cite{Sakai2015,Yanai2014,Gomez2013,Sancho-Parramon2012} and using localized emitters to excite the dark plasmons through evanescent fields \cite{Peyskens2015,Liu2009}. Dark plasmons can also be observed indirectly through their interaction with a bright mode, which generates asymmetric line shapes called Fano resonances\cite{Luk'yanchuk2010,Zhou2011}.
	In our case, the dark plasmon excitation is explained by field retardation. Although the particles ($d=50$~nm) are approximately a tenth of the wavelength of the incident light ($\lambda_{inc}\in[500\mbox{~nm},1000\mbox{~nm}]$ for gold), the bilayer is more than 100~nm thick and the quasi-static approximation is no longer valid. At specific times, the incident electric field points in opposite directions in the two layers, inducing anti-parallel dipole moments in the top and bottom layer. Also the activation of the dark mode is facilitated by the large refractive index of the nanoparticle layers which shrinks the internal wavelength of light inside the metafilm\cite{Kim2018}. Similar dark plasmons were previously observed in studies for metal-insulator-metal (MIM) nanodisks, in which the same retardation effects take place \cite{Wang2017,Chang2012,Frederiksen2013,Verre2015,Zhang2015,Cai2016,Pakizeh2006}. Moreover, Mueller et al. were able to probe by microabsorbance measurements that such plasmons can indeed be excited in bilayers of colloidal nanoparticles\cite{Mueller2018}.
	\condcolor{red}{ We stress also that, due to the rotational symmetry of the lattice, this dark plasmon can be excited with any linearly polarized light.}
		
	The dark mode is observed in both bilayers of gold, Fig. \ref{fig:Bilayer_Default-ARPol}(a), and silver nanoparticles, Fig. \ref{fig:Bilayer_Default-ARPol}(b). 
	\condcolor{red}{ 
	The dark plasmon in silver nanoparticle layers occurs at larger energies than for gold and should even remain in the visible range when the nanoparticles are embedded in a surrounding medium with larger refractive index.}
	For silver, the bright mode at $E^{(Ag)}_B$ can also be studied. Although there is no corresponding peak in the polarization at this energy region, $<P_E>$ has the same sign in both layers, indicating that this mode is bright with the net dipole of each layer pointing in the same direction.
	It is worth emphasizing that the possibility of activating a dark mode with linearly polarized light makes it much easier to use such dark excitations for applications such as SERS. 
	
	\subsection{Diameter Dependence}

			\begin{figure}[!ht]
				\centering
				\includegraphics[page=1]{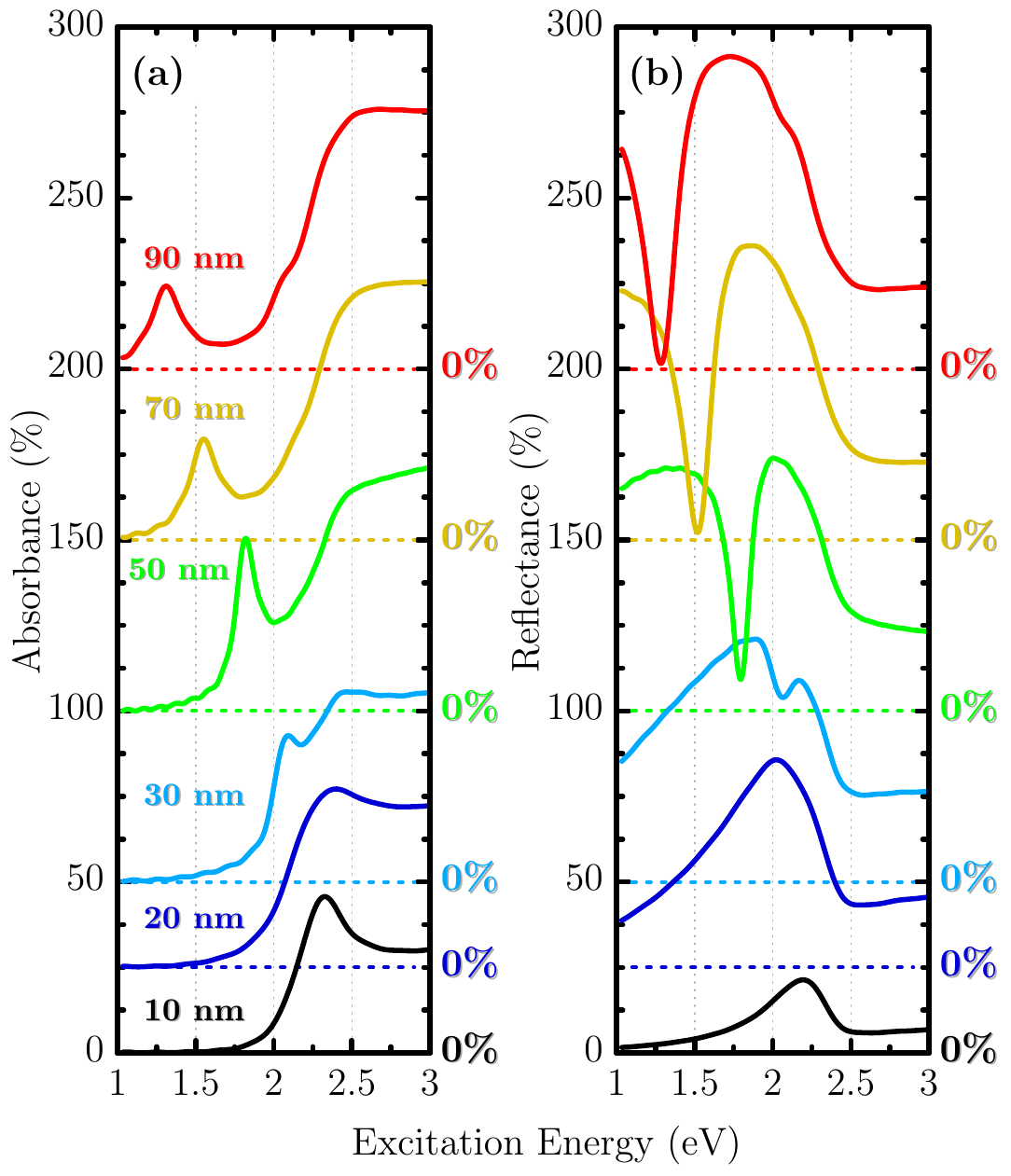}
				\caption{Simulated {\bfseries (a)} absorbance and {\bfseries (b)} reflectance spectra for different gold nanosphere diameters with $g=2$~nm. The color dashed lines indicate the offset of each spectrum. The pronounced absorbance peak related to mode $D$ is strongly redshifted with increasing $d$ followed by its respective reflectance dip.}
				\label{fig:Au_NL2_SwDiam_Waterfall}
			\end{figure}
			
			In the following, we will  evaluate the influence	of the structural parameters on the plasmon resonances of the bilayers, beginning with particles sizes. In Fig. \ref{fig:Au_NL2_SwDiam_Waterfall} the absorbance and reflectance spectra for bilayer systems with different $d$ are presented. The mode $D$ is excited for particles with $d\ge30$~nm. 
			For $d<30$~nm, the retardation effect is weaker and the system gets closer to the quasi-static limit, making it more difficult to excite $D$. We note a redshift in $E^{(Au)}_D$ with increasing $d$, which stems from the fact that the amount of material per unit volume increases, making the coupling between both layers stronger. The anti-parallel dipole configuration of $D$ corresponds to a binding configuration. Therefore, stronger coupling decreases the mode energy, i.e., a redshift. Similar shifts were observed in MIM nanodisk arrays when changing disk size \cite{Wang2017,Chang2012}. There is also a slight increase in $\Gamma^{(Au)}_D$, which stems from the increase in radiative damping. However, the peak integrated area is not significantly affected by the peak broadening, since a proportional decrease in the peaks height is also observed (see Fig. S4 of the Supporting Information). The bright mode is well pronounced for particles with $d<20$~nm but gets damped by the interband transitions for $d>30$~nm. 
			The results for silver are shown in Fig. S5 of the Supporting Information. Similarly to gold, $E^{(Ag)}_D$ also goes towards smaller energies when $d$ increases. The mode $B$ shows a weak redshift for larger $d$ and an increase in absorbance intensity. In addition, as the interband transitions get far away on the energy scale, higher-order modes, such as quadrupoles and hexapoles, are also excited for $d\ge50$~nm.
				
			\begin{figure}
				\centering
				\includegraphics[page=5,trim=0 0 0 180,clip]{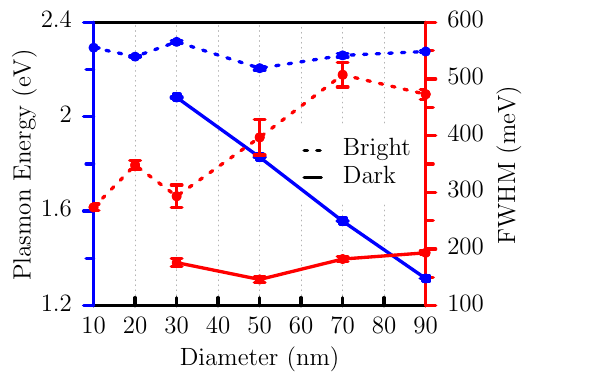}
				\caption{Energy ($E$) and FWHM ($\Gamma$) for the absorbance spectra in Fig. \ref{fig:Au_NL2_SwDiam_Waterfall} as function of the diameter ($d$). The blue lines show the values of spectral position ($E_j$) while the red lines present the FWHM values ($\Gamma_j$). The continuous lines correspond to the results for the dark mode ($D$) and the dashed lines to the bright mode ($B$). The error bar for each value is also provided \condcolor{red}{and some of them may not be visible as they are smaller than the point size}.}
				\label{fig:NL2_SwDiam_Fits}
			\end{figure}
			
			To provide a relation between $d$ and the plasmon energy and lifetime, the absorbance spectra were fitted by
			\begin{equation}
				A(\omega)=\alpha F(\omega)+\sum\limits_{j=1}^{n} \dfrac{I_j\Gamma_j^2}{4(\hbar\omega-E_j)^2+\Gamma_j^2},
				\label{eq:AbsorbanceFitFunc}
			\end{equation}
			where $\hbar\omega\equiv E_{exc}$ is the excitation energy, $n$ is the number of plasmon modes, $I_j$ is the intensity, $E_j$ is the spectral position and $\Gamma_j$ is the FWHM of the j-th plasmon mode. 
			For gold, $F(\omega)$ is the characteristic absorbance $A^{(Au)}_0(\omega)$ of a gold thin film\cite{Lodenquai1991,Kovalenko1999}, in which we considered the dielectric function of gold calculated by the Lorentz-Drude model\cite{LeRuInBook2009-2}, while for silver, $F(\omega)$ is a step-like (sigmoid) function that mimics the absorbance spectrum of a silver thin film.
			$F(\omega)$ is included in order to take the interband transitions into account in the fittings and $\alpha$ is the fitting parameter that accounts for the intensity of these transitions in the material. 
			
			In Fig. \ref{fig:NL2_SwDiam_Fits}, we plot the peak positions and widths of the spectra in Fig. \ref{fig:Au_NL2_SwDiam_Waterfall}(a) as a function of particle diameter. As for $d<30$~nm only the bright mode is present in the spectra, there are no results for the dark mode at this range of diameters.
			As already mentioned, $E^{(Au)}_D$ shifts to smaller energies with increasing $d$, full blue line in Fig. \ref{fig:NL2_SwDiam_Fits}. It follows an approximately linear behavior with 13~meV/nm. Mode $B$, in contrast, remains constant in energy.
			The small apparent oscillations of the bright mode position may be due to the interference with the interband transitions, since they entirely overlap (see Fig. S3 of the Supporting Information), making it hard to capture the real behavior of the mode. 
			 The FWHM is approximately constant for mode $D$ with a small increase for $d>50$~nm (full red line in Fig. \ref{fig:NL2_SwDiam_Fits}), while mode $B$ becomes broader for larger spheres because radiative damping scales with sphere volume (dashed red line). 

	\subsection{Influence of Lattice and Stacking}
		
		\begin{figure}[t]
			\centering
			\includegraphics[trim=0 0 95 30,clip]{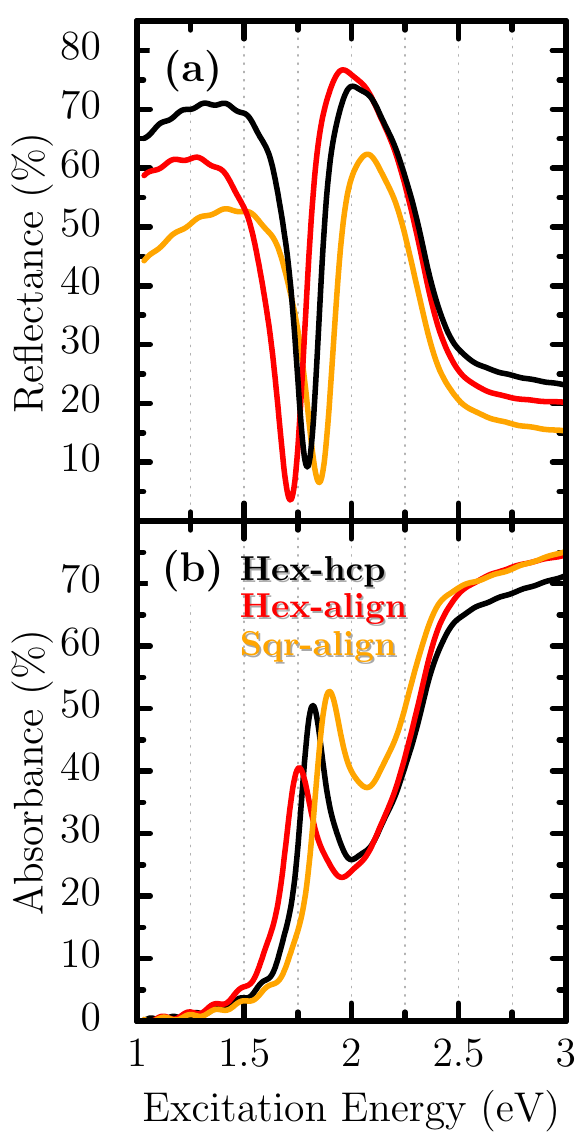}
			\caption{Influence of lattice and stacking on the {\bfseries (a)} reflectance and {\bfseries (b)} absorbance spectra. Three different configurations were studied: a hexagonal lattice with the layers in a hexagonal close-packed stacking (black lines) and aligned on top of each other (red lines) and a square lattice with the particles aligned over each other (orange lines). In all cases, $d=50$~nm and $g=2$~nm were used as parameters.}
			\label{fig:NL2_SwLatStk}
		\end{figure}
		We now investigate the effect of the lattice and the stacking of the layers on the plasmonic properties of the bilayers. Figure \ref{fig:NL2_SwLatStk} presents the optical response for bilayers with three different particle arrangements: layers formed by a hexagonal lattice with a hexagonal close-packed stacking (Hex-hcp) and with the particles aligned along $z$ (Hex-align) and layers formed by a square lattice with aligned stacking (Sqr-align). The main features of the spectra are present in all three configurations. $E^{(Au)}_D$ is redshifted by 64~meV for Hex-align and blueshifted by 82~meV for Sqr-align with respect to Hex-hcp. This can be understood from the inset of Fig. \ref{fig:Bilayer_Default-ARPol}(d). In the Hex-hcp stacking, the positive (negative) charges in the particle are closer to each other than in Hex-align, which causes a larger repulsion and blueshift. The blueshift for Sqr-align stacking stems from the lower particle density in this configuration compared to Hex-align. The smaller density results in less coupling and the plasmon energy has a smaller redshift than in the Hex-align configuration. The reduced coupling induced from the in-plane lattice configuration has a stronger effect than the layers stacking, as the blueshift in the Sqr-align configuration (compared to the Hex-align) is larger than the blueshift for Hex-hcp.

	\subsection{Gap Dependence}

		\begin{figure}[!ht]
			\centering
			\includegraphics[page=1]{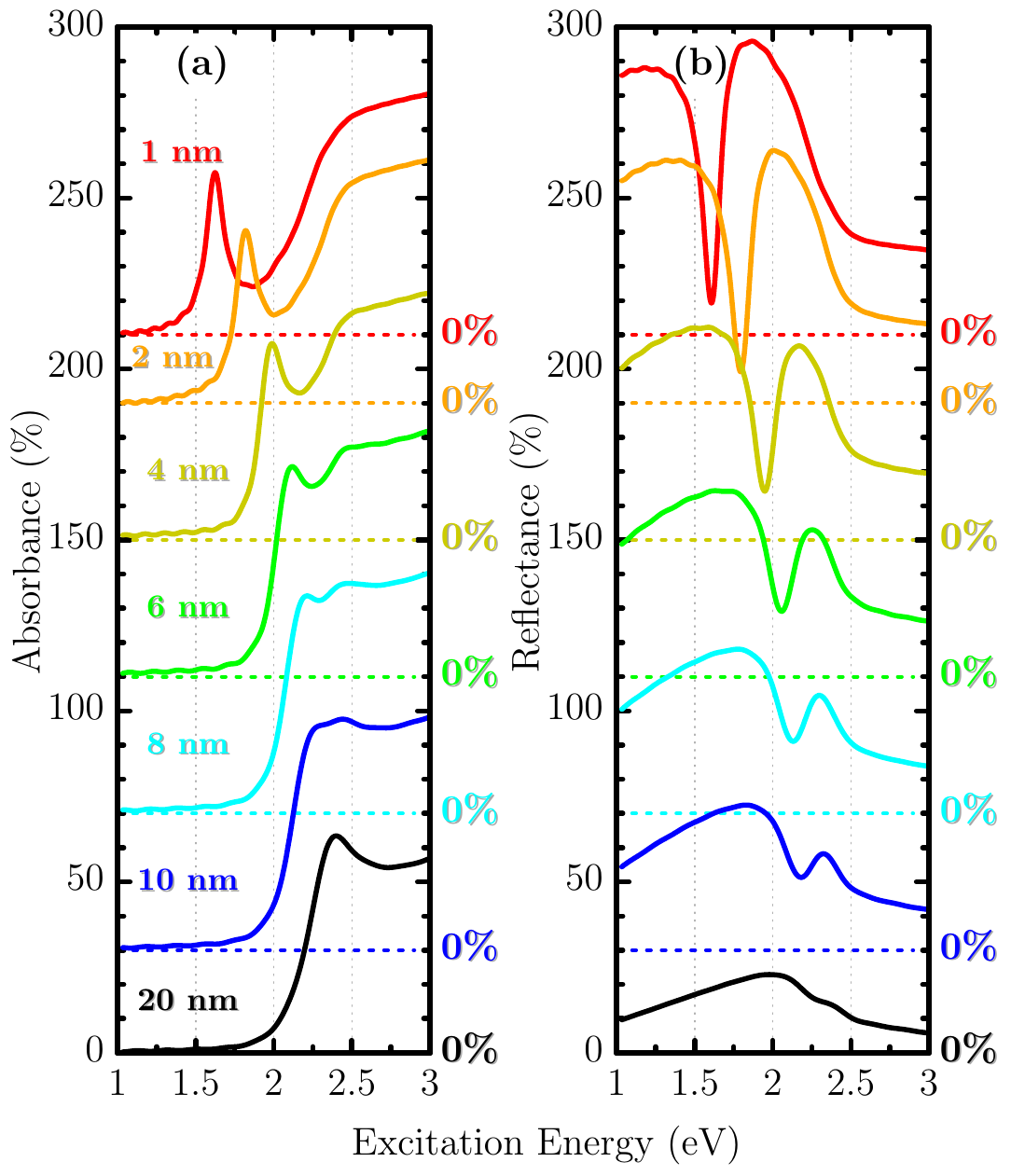}
			\caption{{\bfseries (a)} Absorbance and {\bfseries (b)} reflectance spectra for different gap sizes between all gold particles ($g$) with $d=50$~nm. The color dashed lines indicate the offset of each spectrum. It can be seen that the dark mode is effectively redshifted with decreasing $g$.}
			\label{fig:Au_NL2_SwAllGap_Waterfall}
		\end{figure}
		
\begin{figure}
			\centering
			\includegraphics[page=2,trim=0 0 35 180,clip]{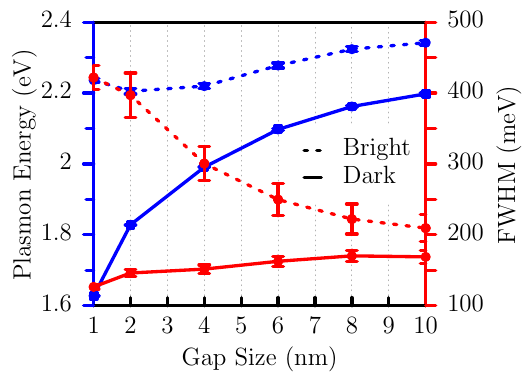}
			\caption{Fitted parameters for the absorbance spectra provided in Fig. \ref{fig:Au_NL2_SwAllGap_Waterfall} as function of the gap size ($g$). The blue lines show the values of spectral position ($E_j$) while the red lines present the FWHM values ($\Gamma_j$). The continuous lines correspond to the results for mode $D$ and the dashed lines to mode $B$. The error bar for each value is also provided \condcolor{red}{and some of them may not be visible as they are smaller than the point size}.}
			\label{fig:NL2_SwAllGap_Fits}
		\end{figure}

		We now turn our attention to the influence of the gap size $g$ ($g\equiv g_p=g_l$) on the optical properties. In Fig. \ref{fig:Au_NL2_SwAllGap_Waterfall} the absorbance and reflectance spectra for systems with varying $g$ are shown. The dark mode $D$ starts to be effectively activated for $g\le10$~nm.
		As $g$ is further reduced, $E^{(Au)}_D$ shifts towards smaller energies due to stronger coupling. The bright mode $B$ shows a similar behavior with shrinking $g$ as that observed for increasing $d$, i.e., it becomes completely suppressed by the interband transitions.
		Similar calculations for silver (Fig. S6 of the Supporting Information) show qualitatively the same effect. Remarkably, even for $g=20$~nm, the modes $B$ and $D$ of silver can still be distinguished, which does not occur for gold.
		By fitting the calculated spectra similar to the diameter analysis, the strong redshift of $E^{(Au)}_D$ with decreasing $g$ becomes apparent.
		Figure \ref{fig:NL2_SwAllGap_Fits} shows that $E^{(Au)}_D$ (full blue line) goes from 2.2 to 1.6~eV when $g$ changes from 10 to 1~nm. $\Gamma^{(Au)}_D$ (full red line) is reduced when $g$ decreases while $\Gamma^{(Au)}_B$ (dashed red line) increases. 
		These results reveal that the energy of $D$ can be tuned over a broad spectral range in the visible/near-infrared by changing the gap size. 
		Colloidal samples in which $g$ spreads over many values are expected to show $D$ modes with broader FWHM than that of perfect NP crystals (simulated layers) as indeed observed experimentally\cite{Mueller2018}.

		\begin{figure}[!ht]
			\centering
			\begin{tikzpicture}
				\node (plot) {\includegraphics{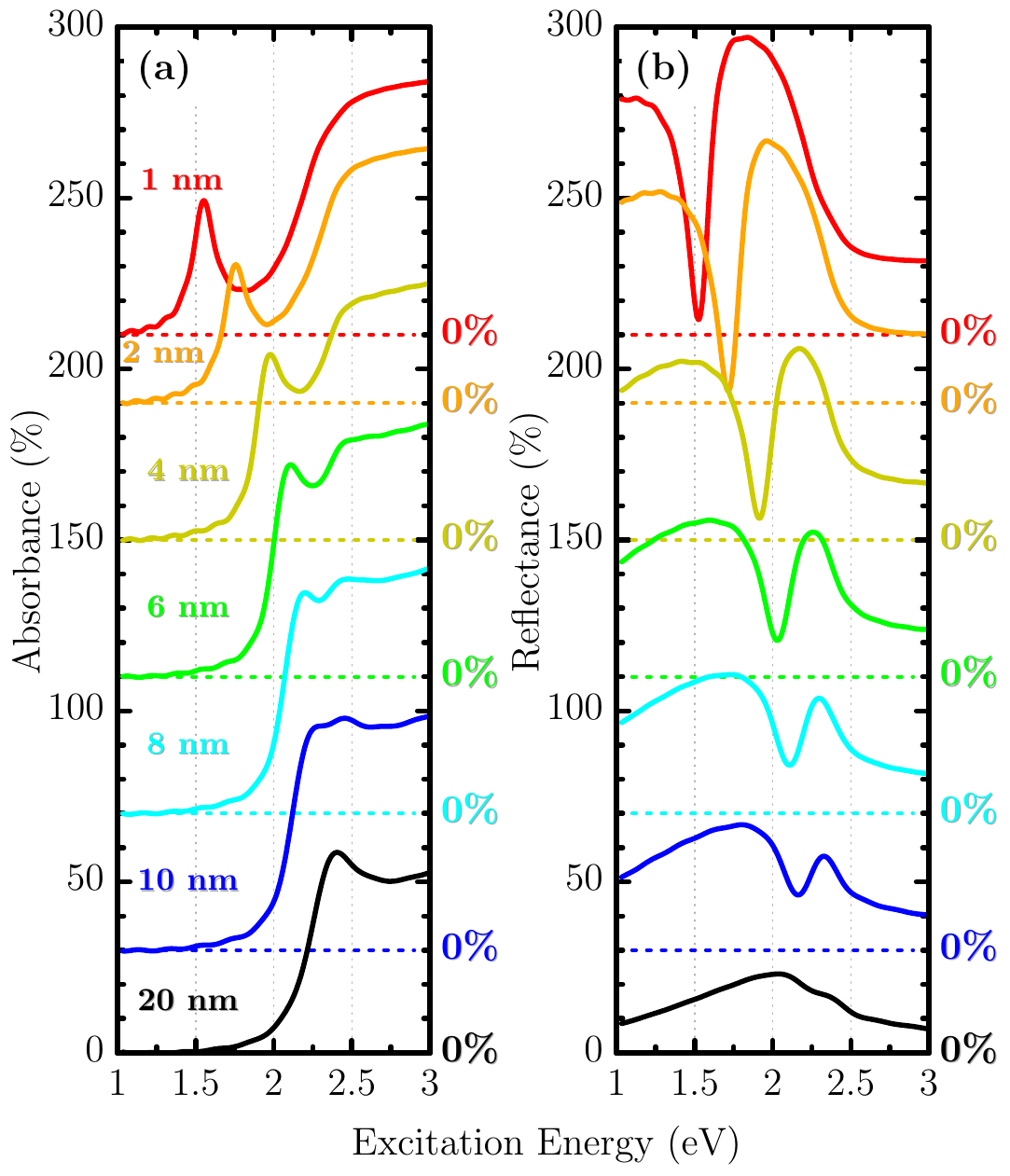}};
				\node[below right] at ([shift={(-.95,-.3)}]plot.north east)
				{\includegraphics[width=.135\paperwidth,trim=900 100 900 100,clip]{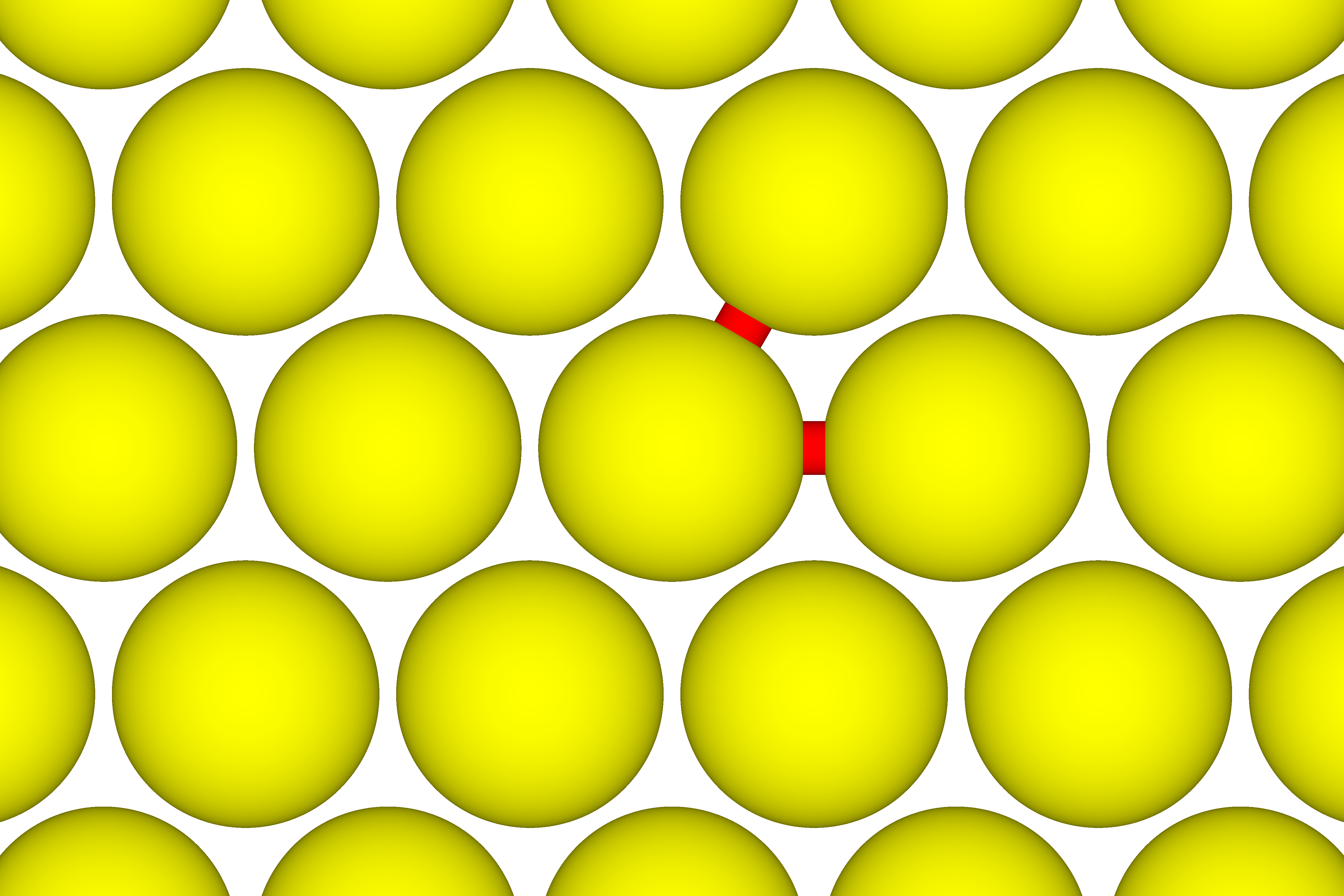}};
			\end{tikzpicture}
			\caption{{\bfseries (a)} Absorbance and {\bfseries (b)} reflectance spectra for different intralayer gaps $g_p$. The gold bilayers have $d=50$~nm and $g_l=2$~nm. Aligned layer stacking was used. The color dashed lines indicate the offset of each spectrum and $g_p$ is represented by the small red lines in rightmost illustration.
			}
			\label{fig:NL2_SwIntraGap_Waterfall}
		\end{figure}
		
		We now investigate the influence of $g_p$ and $g_l$ separately. We consider hexagonal bilayers with aligned stacking (Hex-align), as $g_p$ and $g_l$ are linearly independent in this structure. Figure \ref{fig:NL2_SwIntraGap_Waterfall} shows that changes in the intralayer gap $g_p$ affect the optical properties in a similar fashion as when varying both gaps simultaneously. A strong redshift of $D$ is verified when $g_p$ is reduced, which indicates that this mode is highly sensitive to it. Also, for $g_p>10$~nm, this mode enters the interband transitions region and can no longer be identified. 
				
		\begin{figure}[!ht]
			\centering
			\begin{tikzpicture}
				\node (plot) {\includegraphics{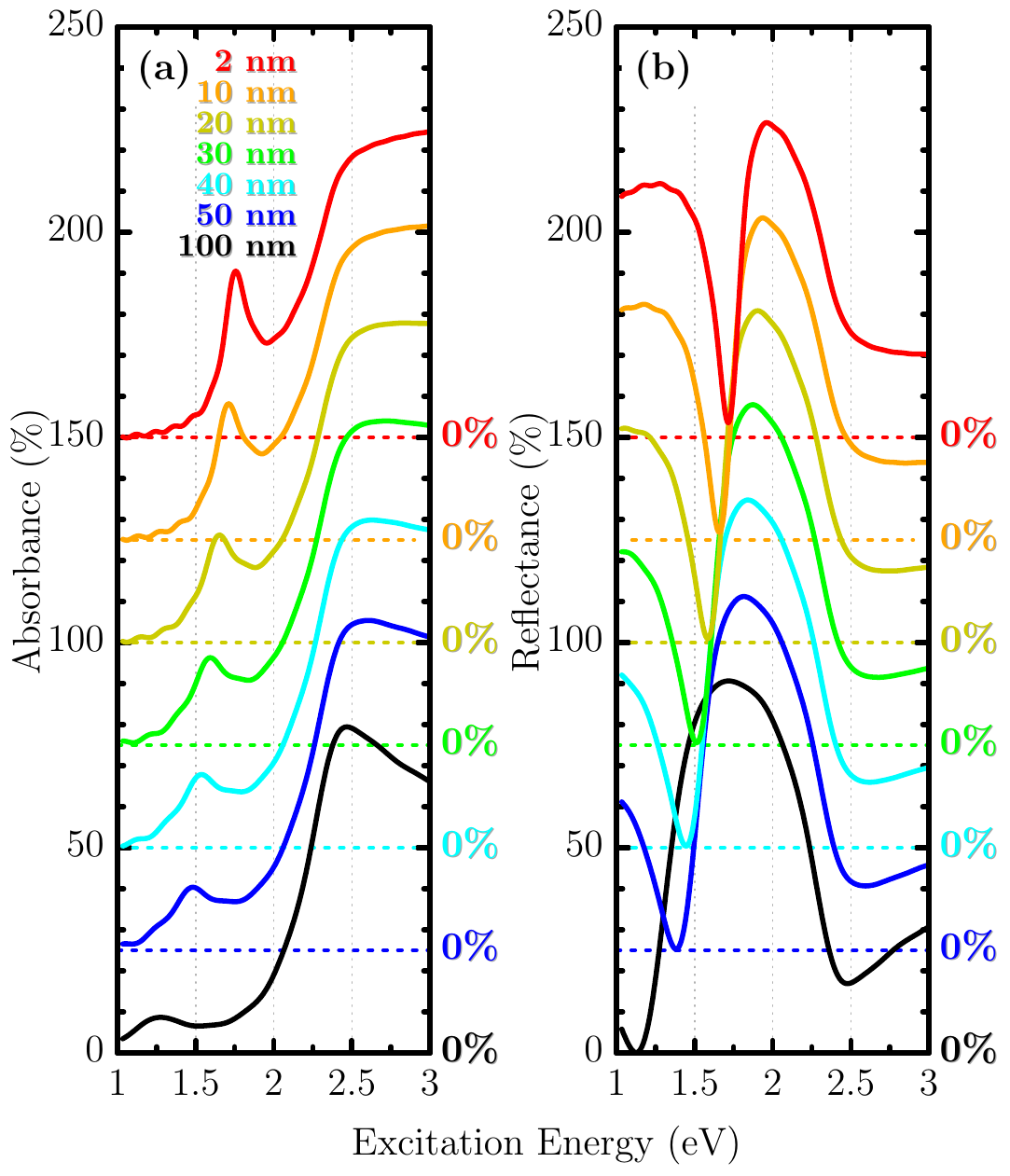}};
				\node[below right] at ([shift={(-.95,-.3)}]plot.north east)
					{\includegraphics[width=.12\paperwidth,trim=0 400 820 00,clip]{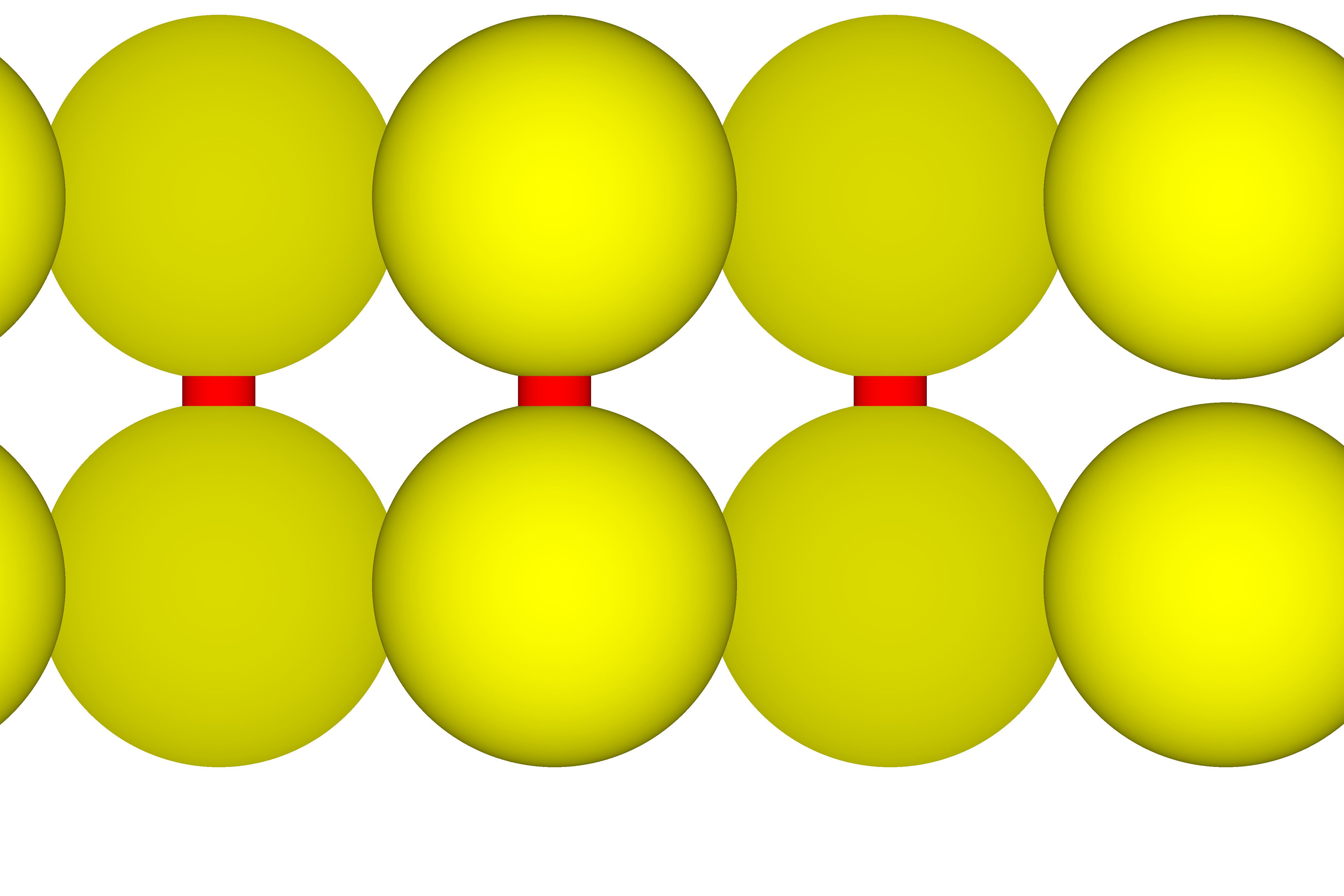}};
			\end{tikzpicture}
			\caption{{\bfseries (a)} Absorbance and {\bfseries (b)} reflectance spectra for gold bilayers with different $g_l$ ($d=50$~nm, $g_p=2$~nm, aligned stacking). The color dashed lines indicate the offset of each spectrum and $g_l$ is presented as the small red lines in the rightmost panel.}
			\label{fig:NL2_SwInterGap_Waterfall}
		\end{figure}
		Regarding the interlayer gap $g_l$ dependence, Fig. \ref{fig:NL2_SwInterGap_Waterfall} shows $D$ already in the near-infrared ($E^{(Au)}_D=1.28$~eV) for $g_l=100$~nm and that it blueshifts with decreasing $g_l$. Note that the range in which $g_l$ is varied ($2-100$~nm) is much larger than that used for $g_p$ ($1-20$~nm), as, otherwise, no change would be observed. 
		This fact suggests a low sensitivity of $E^{(Au)}_D$ on $g_l$, in contrast to the high sensitivity on $g_p$. This difference is explained by verifying that, for each nanoparticle and at any given distance, there are always more neighbors from the same layer (related to $g_p$) than from the other layer (related to $g_l$). For example, for the Hex-align configuration, each nanoparticle has six first neighbors within the same layer while there is just one from the other layer.
		Regardless of the magnitude of this effect, it appears surprising to observe a blueshift when $g_l$ decreases considering the dipole hybridization model. One might expect the inverse in layer-layer coupling and a redshift with decreasing $g_l$. 
		
		\begin{figure}
			\centering
			\def\h{5cm}
			\begin{tikzpicture}
				\matrix[column sep=0mm,matrix of nodes,ampersand replacement=\&,nodes={inner sep=0mm}] (figs) {
				\includegraphics[page=1,height=\h,trim=10 0 50 40,clip]{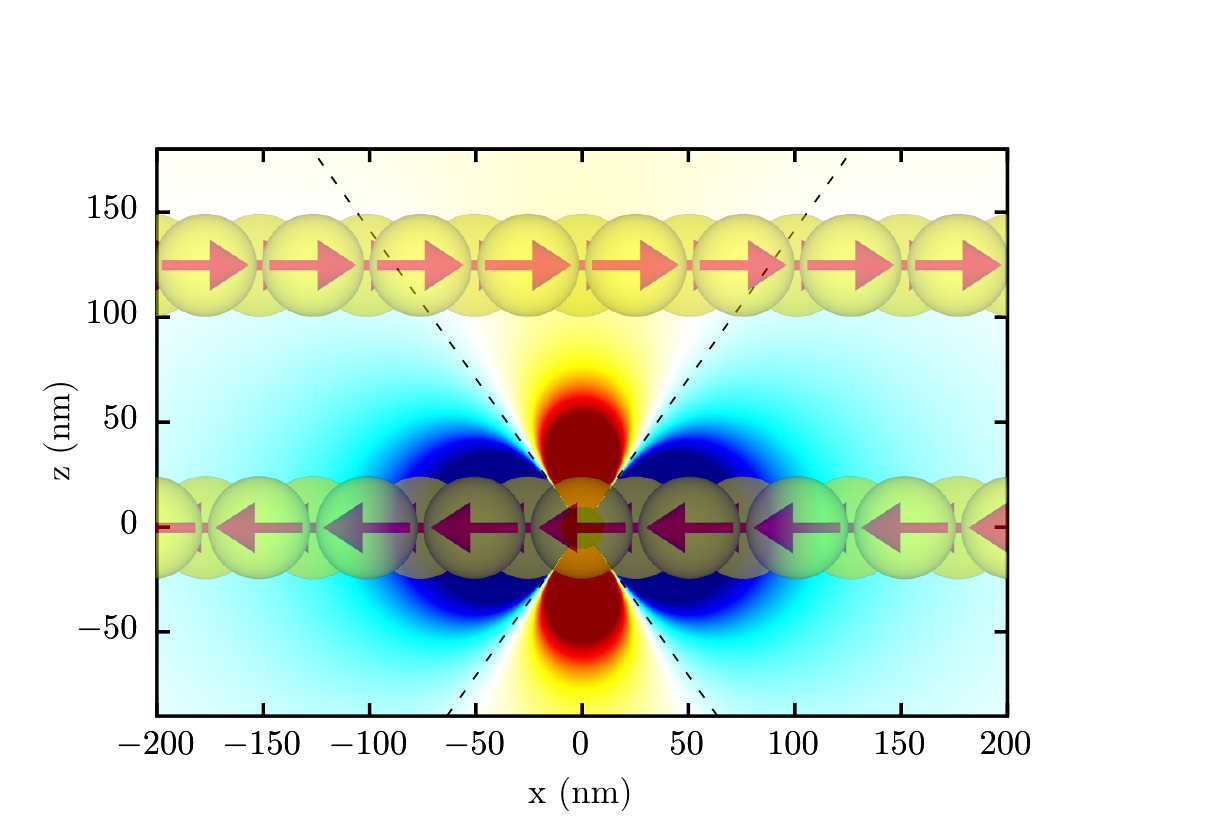}\&
				\includegraphics[page=2,height=\h,trim=10 0 0 40,clip]{Figs/DipDip_Interaction_Illust}\\};
				\foreach \i/\text in {1/(a),2/(b)}\node[font=\bfseries,above right] at (figs-1-\i.north west) {\text};
			\end{tikzpicture}
			\caption{Mapping of the real part of the dipole-dipole interaction $W_{ij}$ between a central dipole and {\bfseries (a)} all dipoles in the other layer and {\bfseries (b)} all other dipoles in the same layer. Panel {\bfseries (a)} shows the side view of a bilayer in the dipole configuration of mode $D$ and panel {\bfseries (b)} the top view of a monolayer in mode $B$ configuration. An illustration of the nanoparticles is superimposed to the mappings for a better comprehension. The arrows represent the dipole moment in each particle and the dashed lines separate the regions where the dipole-dipole interaction has opposite signs.}
			\label{fig:DipDipInt}
		\end{figure}
		In order to understand the observed behavior, we take a closer look at the interaction between two electric dipoles of the bilayer lattice, which is expressed by the following potential energy\cite{GreinerInBook1998}
		\begin{equation}
				W_{ij}=\dfrac{e^{i\omega R_{ij}/c}}{4\pi\epsilon_0}\vect{p}_{i}\cdot\vect{p}_{j}\left[\left(\dfrac{1}{R^3_{ij}}-\dfrac{i\omega}{c}\dfrac{1}{R^2_{ij}}\right)\left(1-3\cos^2\phi_{ij}\right)-\dfrac{\omega^ 2}{c^2R_{ij}}\sin^2\phi_{ij}\right]
			\label{eq:DipDipEnergy}
		\end{equation}
		where $\vect{p}_{i}$ is the dipole moment of the particle $i$, $R_{ij}\equiv|\vect{R}_{ij}|$ is the distance between particles $i$ and $j$, $\phi_{ij}$ is the angle between $\vect{R}_{ij}$ and the polarization direction of the incident field, $\omega$ is the angular frequency of the incident field, $c$ is the light speed in vacuum, and $\epsilon_0$ is the vacuum permittivity. From Eq. \eqref{eq:DipDipEnergy}, it is clear that $W_{ij}$ can be either positive or negative depending on $\phi_{ij}$. 
		In Fig. \ref{fig:DipDipInt}(a), we illustrate that, for mode $D$  and a given particle $i$ in one layer, there are regions in which the dipole-dipole interaction is attractive (red regions) and others in which it is repulsive (blue regions). These regions are separated by the two dashed lines in the figure, which represent two cones in 3D space and indicate the exact angle $\phi_0$ in which $W_{ij}$
		changes sign. Considering the effect of decreasing $g_l$, more particles get inside the repulsive regions and the overall interaction between the layers becomes more repulsive. This means that the mode moves towards higher energies, thus explaining the observed blueshift.
		
		It is worth noting that previous studies on similar systems observed a redshift when $g_l$ diminishes, e.g., MIM vertical dimers\cite{Ogier2016,Frederiksen2013}, MIM pentamers\cite{Verre2015} and MIM nanodisk arrays\cite{Chang2012}. The apparent contradiction to our results is explained by the fact that the distance between adjacent MIM nanoparticles (corresponding to $g_p$) is much larger in Refs. \citenum{Chang2012}, \citenum{Frederiksen2013}, \citenum{Verre2015} and \citenum{Ogier2016}  than the $g_p$ considered here. 
		Consequently, the coupling between these MIM NPs is not strong enough for a similar blueshift to be verified. The blueshift occurs when $g_p$ is sufficiently small for NPs to be able to move from the attractive region to the repulsive region when $g_l$ is reduced. 
		We can estimate the minimum $g_p$ necessary for a redshift to occur when the layer distance is decreasing starting from a given height. It turns out that this minimum $g_p$ needs to be very large for a possible redshift be observed. In Fig. \ref{fig:NL2_SwInterGap_Waterfall} for example, $g_p$ needs to be greater than $130$~nm for a redshift to be observed starting from a $g_l$ of 50~nm. At this regime, the particles are too far for the dark mode being effectively excited (as observed in Fig. \ref{fig:NL2_SwIntraGap_Waterfall}), meaning that the previously reported redshifts are not expected to appear in the layers of metallic nanospheres.
		
		
		Equation \eqref{eq:DipDipEnergy} may also be used to understand why no strong redshift occurs for mode $B$ in the monolayer when $g$ is reduced. In this case, there are also regions of attractive and repulsive interactions, see Fig. \ref{fig:DipDipInt}(b). However the gap size does not significantly affect the number of particles in each region, which means that the redshift is not related to a change in this number.
		The slight redshift in Fig. \ref{fig:Au_D10NL1_SwGap} is caused by the binding coupling between the particles in the red regions of Fig. \ref{fig:DipDipInt}(b), which is slightly stronger than the anti-binding coupling between the particles in the blue regions. Similar explanations apply to the redshifts observed for the bright mode in silver with decreasing $g$ and increasing $d$.
				
	\subsection{Nearfield}
			
		\begin{figure}[!ht]
			\centering
			\begin{tikzpicture}[global scale=.9]
				\def\h{4cm}
				\matrix[column sep=0mm,matrix of nodes,ampersand replacement=\&,nodes={inner sep=0mm}] (plot) {
					\node (plot-1-1) {\includegraphics[page=1,trim=0 0 0 15,clip]{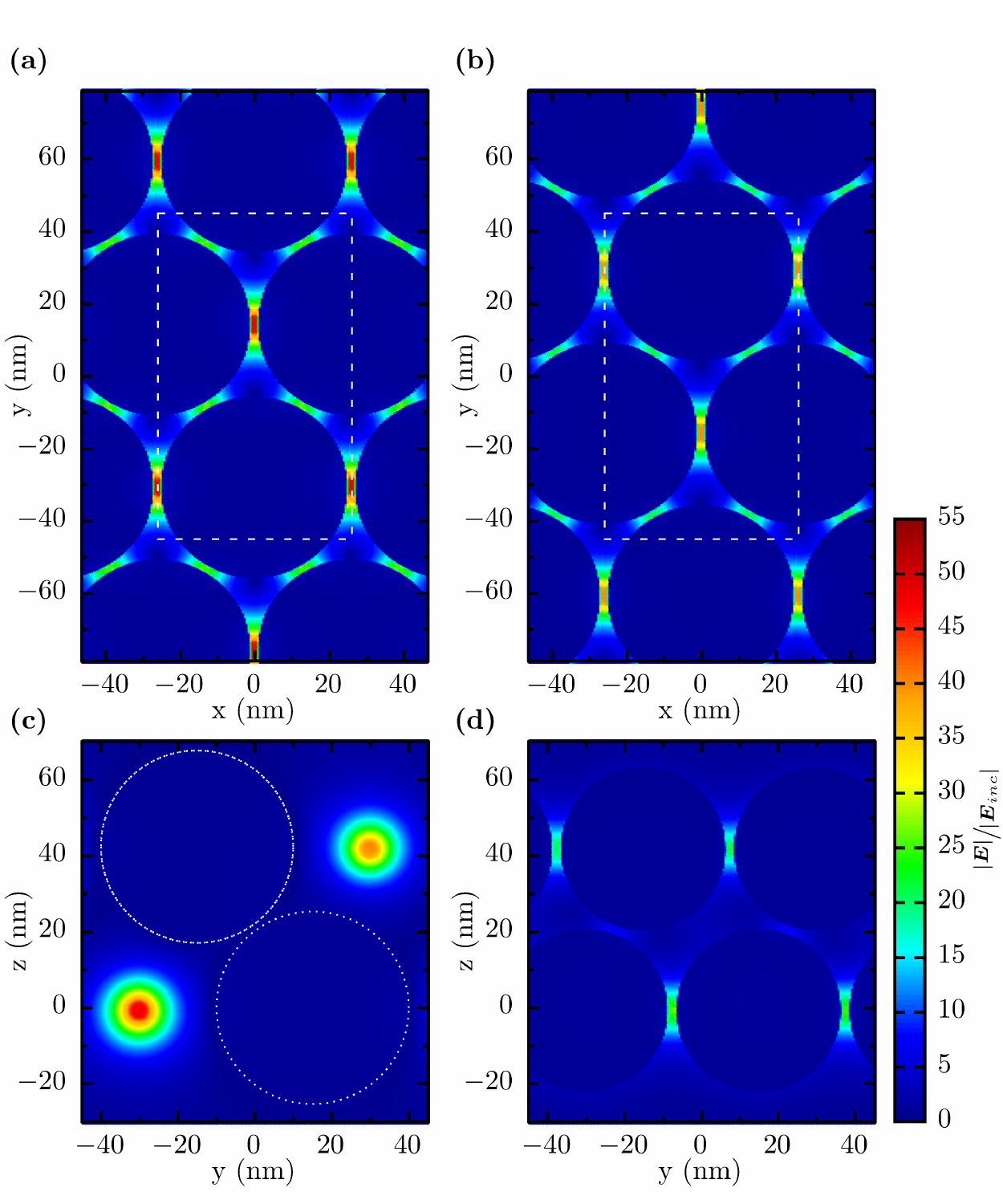}};\\
				};

				\node[inner sep=0,right] (illust)
						at ([shift={(-1.4,4.8)}]plot.east) 
						{\includegraphics[width=.18\paperwidth,trim=00 0 00 0,clip]{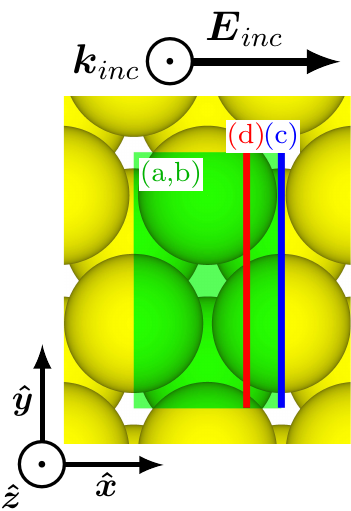}};
			\end{tikzpicture}
			\caption{Nearfield enhancement factor mapping for the gold bilayer of Fig. \ref{fig:Bilayer_Default-ARPol} at the excitation energy of mode $D$ ($E_{exc}=E^{(Au)}_D=1.82$~eV). The two upper panels correspond to $xy$ planes connecting the centers of each sphere in the {\bfseries (a)} first layer and {\bfseries (b)} second layer. The unit cells of each respective layer are indicated by the white dashed lines at the panels and by the green area shown in the inset. The two lower panels are related to the $yz$ planes illustrated in the inset by {\bfseries (c)} the blue line and {\bfseries (d)} the red line.
			}
			\label{fig:Bilayer_Default-Efield}
		\end{figure}

		We finally turn our attention to the nearfield characteristics of the bilayers. Figure \ref{fig:Bilayer_Default-Efield} displays the enhancement factor $EF=\frac{|\vect{E}|}{|\vect{E}_{inc}|}$ of the nearfield amplitude at the most relevant regions of the gold bilayer with $d=50$~nm and $g=2$~nm, i.e., the system analyzed in Fig. \ref{fig:Bilayer_Default-ARPol}. At $E_{exc}=E^{(Au)}_D$, strong enhancement factors of up to $EF=55$ are observed in the regions around the particle gaps. 
		\condcolor{red}{
		Such an $EF$ reinforces the potential of these layers for SERS applications, since an SERS enhancement of up to $\left(\frac{|\vect{E}|}{|\vect{E}_{inc}|}\right)^4\approx10^7$ is expected. Even though this SERS enhancement is lower than for a nanoparticle dimer, the layers have a larger density of hotspots that can provide enhancement. While a dimer has just one hotspot, the bilayers have approximately $10^3$ hotspots per $\mu\mbox{m}^2$. Taking into account the tunability with particles size and gap, the potential of this mode and, consequently, of these materials, for applications in surface-enhanced spectroscopy becomes evident.}
		
		Remarkably, there are no significant hotspots between the layers. The hotspots are found at the $g_p$ voids such that the most intense places are for those oriented parallel to the incident light. The absence of enhancement between the layers is explained by the orientation of all dipoles within the layers, as the dipoles have no component along the $z$ direction. 
		Our nearfield results are in good agreement with other mappings previously reported in the literature\cite{Cai2016,Alaeian2012}.

\section{Conclusion \label{sec:conc}}

	In conclusion, we investigated the plasmonic properties of mono- and bilayers made of gold and silver nanoparticles when illuminated by linearly polarized light at normal incidence. We demonstrated by FDTD simulations the excitation of two plasmon modes in the bilayers: a bright mode, characterized by a non-zero net dipole moment, and a dark mode that has a vanishing dipole moment. The dark mode was excited due to field retardation, inducing an anti-parallel dipole configuration in the bilayers. We also verified that the resonance energy and the linewidth of the dark mode can be tuned over a wide range of energies by varying the structural parameters of the system, such as particles size and their separation. For instance, its resonance energy is redshifted by more than 360~meV when changing the gap between the particles from 4 to 1~nm. 
	\condcolor{red}{ 
	We verified that gold and silver have qualitatively similar optical responses. Besides the differences in the plasmon energy and linewidth of their dark modes, the bright mode of silver bilayers could be visualized in the spectra as it is not completely damped by the interband transitions. For gold, this bright mode is totally suppressed and is not visible in most of the presented spectra.} 
	\condcolor{red}{ 
	Nearfield hotspots where the field amplitude is enhanced by 55 times were observed. This enhancement is ideal for SERS as the nanoparticle layers provide not only SERS enhancement factors on order of $10^7$ but also have $10^3$ hotspots per $\mu\mbox{m}^2$.} 
	\condcolor{red}{ 
	Beyond SERS, the narrow peak width of the dark mode is also ideal for refractive index sensing. 
	Recently, we showed that such plasmons can be excellent channels for generating hot-electrons that can be used for photocatalytical applications\cite{Mueller2018a}.}
	All the discussions presented in this study, together with the already well developed methods for the materials synthesis, show how promising they are for plasmonic applications, being also very intriguing from the theoretical point of view.

\begin{suppinfo}
		\condcolor{red}{ 
		Reflectance and absorbance spectra of silver monolayers ($d=10$~nm) for the same gap sizes ($g$) as used in Fig. \ref{fig:Au_D10NL1_SwGap}. Transmittance spectra of the corresponding bilayers of Fig. \ref{fig:Bilayer_Default-ARPol}. Details on the curve fitting analysis and additional fitting parameters plots. Spectra of silver bilayers for diameter and gap size analysis, analogous to the analysis for gold in Figs. \ref{fig:Au_NL2_SwDiam_Waterfall} and \ref{fig:Au_NL2_SwAllGap_Waterfall}.}
\end{suppinfo}
	
\begin{acknowledgement}
	This work was supported by the European Research Council under grant DarkSERS (772108).
	B.G.M.V. acknowledges the Coordenação de Aperfeiçoamento de Pessoal de Nivel Superior (CAPES) for the financial support under the program PDSE (Grant No. \linebreak 88881.134611/2016-01) and Dahlem Research School (DRS).
	N.S.M. acknowledges financial support from Deutsche Telekom Stiftung.
	E.B.B. and B.G.M.V acknowledge financial support from CNPq, CAPES (finance code 001) and FUNCAP (PRONEX PR2-0101-00006.01.00/15).
	S.R. acknowledges the FocusArea NanoScale.
\end{acknowledgement}


\bibliography{Paper}

\end{document}


This document is the unedited Author's version of a Submitted Work that was subsequently accepted for publication in The Journal of Physical Chemistry C, copyright \textcopyright American Chemical Society after peer review. To access the final edited and published work see the following link: \href{https://pubs.acs.org/doi/10.1021/acs.jpcc.9b03859}{https://pubs.acs.org/doi/10.1021/acs.jpcc.9b03859}.

\newpage

\begin{figure}
	\centering
	\includegraphics[scale=.8,trim=0 0 70 0,clip]{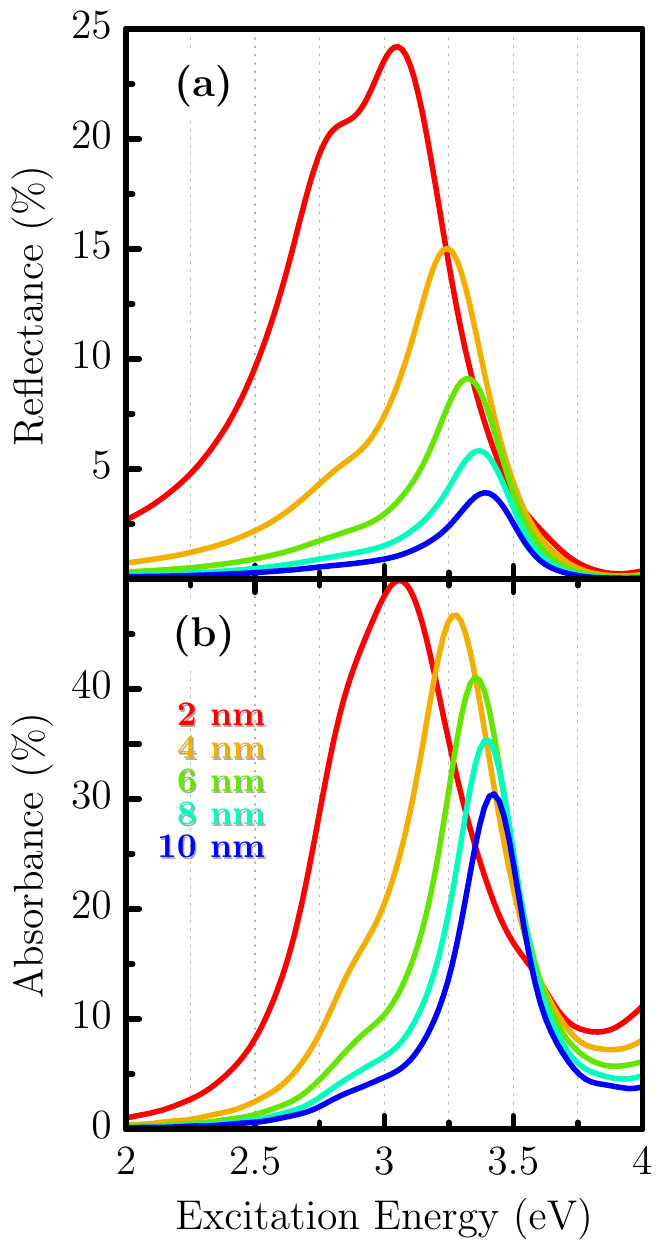}
	\caption{{\bfseries (a)} Reflectance and {\bfseries (b)} absorbance spectra for hexagonal silver monolayers with $d=10$~nm as a function of the gap between the particles.}
	\label{fig:Ag_D10NL1_SwGap}
\end{figure}

\begin{figure}
	\centering
	\includegraphics[page=2,width=.6\textwidth,trim=0 0 0 160,clip]{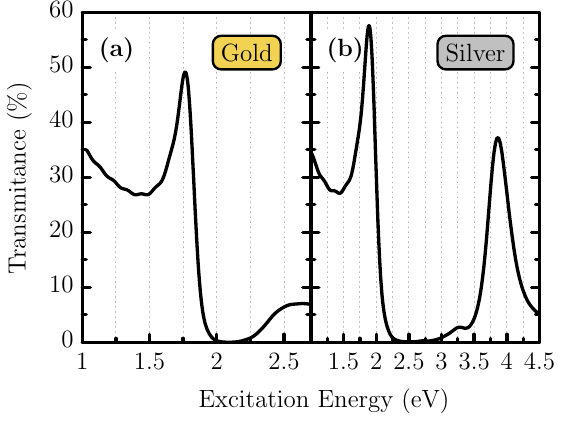}
	\caption{Transmittance spectra of the bilayers of Fig. 5 ($d=50$~nm, $g=2$~nm, hcp stacking). The results for both {\bfseries (a)} gold and {\bfseries (b)} silver are provided.}
	\label{fig:Bilayer_Default-T}
\end{figure}

\begin{figure}
	\centering
	\includegraphics[page=3,height=8cm,trim=0 0 0 155,clip]{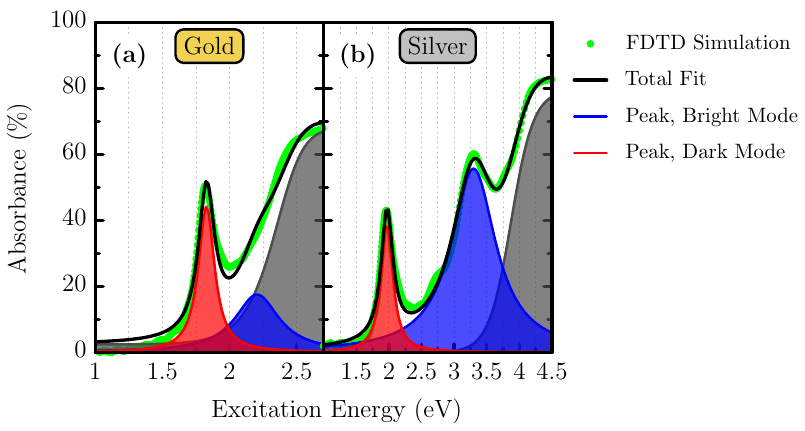}
	\caption{Absorbance spectrum of the bilayers (green curve) in which $d=50$~nm and $g=2$~nm , i.e, the same systems as in Fig. 5. The spectrum is fitted with equation (1) (the black curve), in which two Lorentzian peaks were used. In both panels, the red shaded curve corresponds to the peak of the dark plasmon mode while the blue shaded curve corresponds to the peak of the bright plasmon mode. For gold, the dark mode Lorentzian peak is centered at 1.82~eV and has a FWHM of 146~meV while the bright mode peak is at 2.21~eV and has a FWHM of 396~meV. For silver, the Lorentzian peak of the dark mode is at 1.97~eV and has a FWHM of 217~meV while the bright mode peak is centered at 3.30~eV and has a FWHM of 830~meV. The grey line and area represent the $\alpha F(\omega)$ term in equation (1), which is the term that takes into account the interband transitions of {\bfseries (a)} gold and {\bfseries (b)} silver.
	}
	\label{fig:Bilayer_Default-FitSpecs}
\end{figure}

\begin{figure}
	\centering
	\includegraphics[page=6,width=.6\textwidth,trim=0 0 0 160,clip]{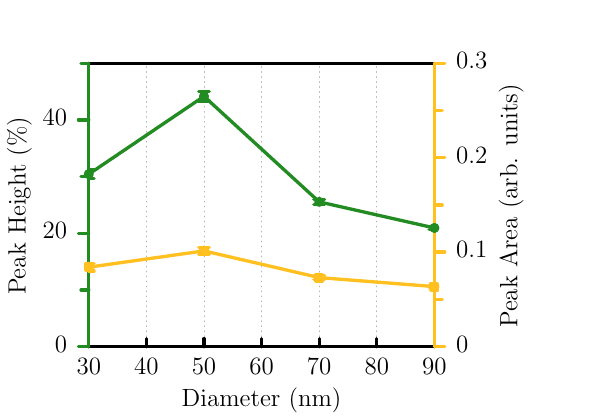}
	\caption{Height ($I$) and area ($A=\pi I\Gamma$) of the dark mode peaks in the absorbance spectra of Fig. 6 as function of the diameter ($d$). The green line shows the peak height of the dark mode ($I^{(Au)}_D$) while the yellow line presents the area of the peak ($A^{(Au)}_D$).
	}
	\label{fig:NL2_SwDiam_FitsHeightArea}
\end{figure}

\begin{figure}
	\centering
	\includegraphics[page=1]{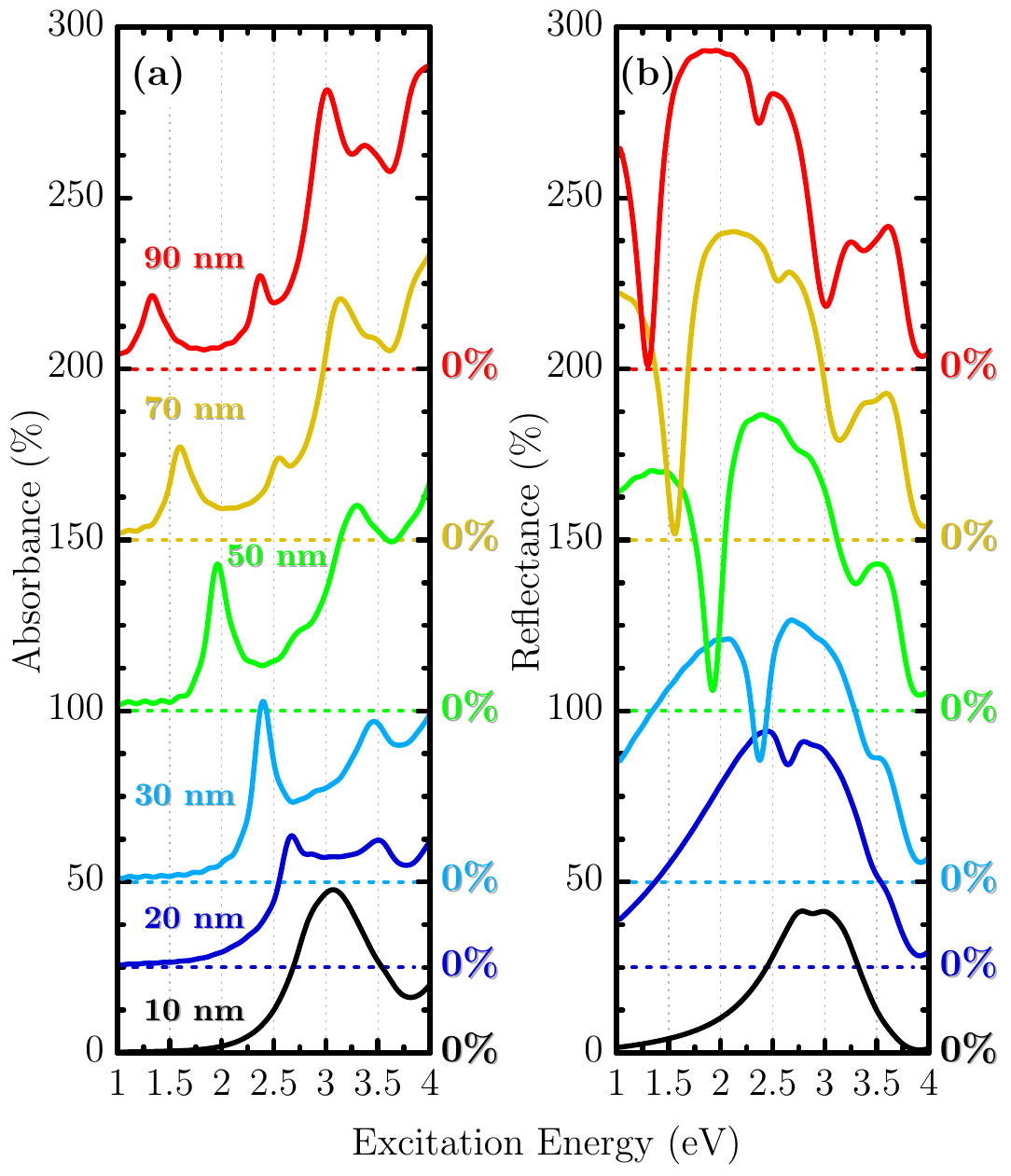}
	\caption{Simulated {\bfseries (a)} absorbance and {\bfseries (b)} reflectance spectra for silver nanospheres with different diameter values. The gaps were fixed to 2~nm. The color dashed lines indicate the offset of each spectrum.}
	\label{fig:Ag_NL2_SwDiam_Waterfall}
\end{figure}

\begin{figure}
	\centering
	\includegraphics{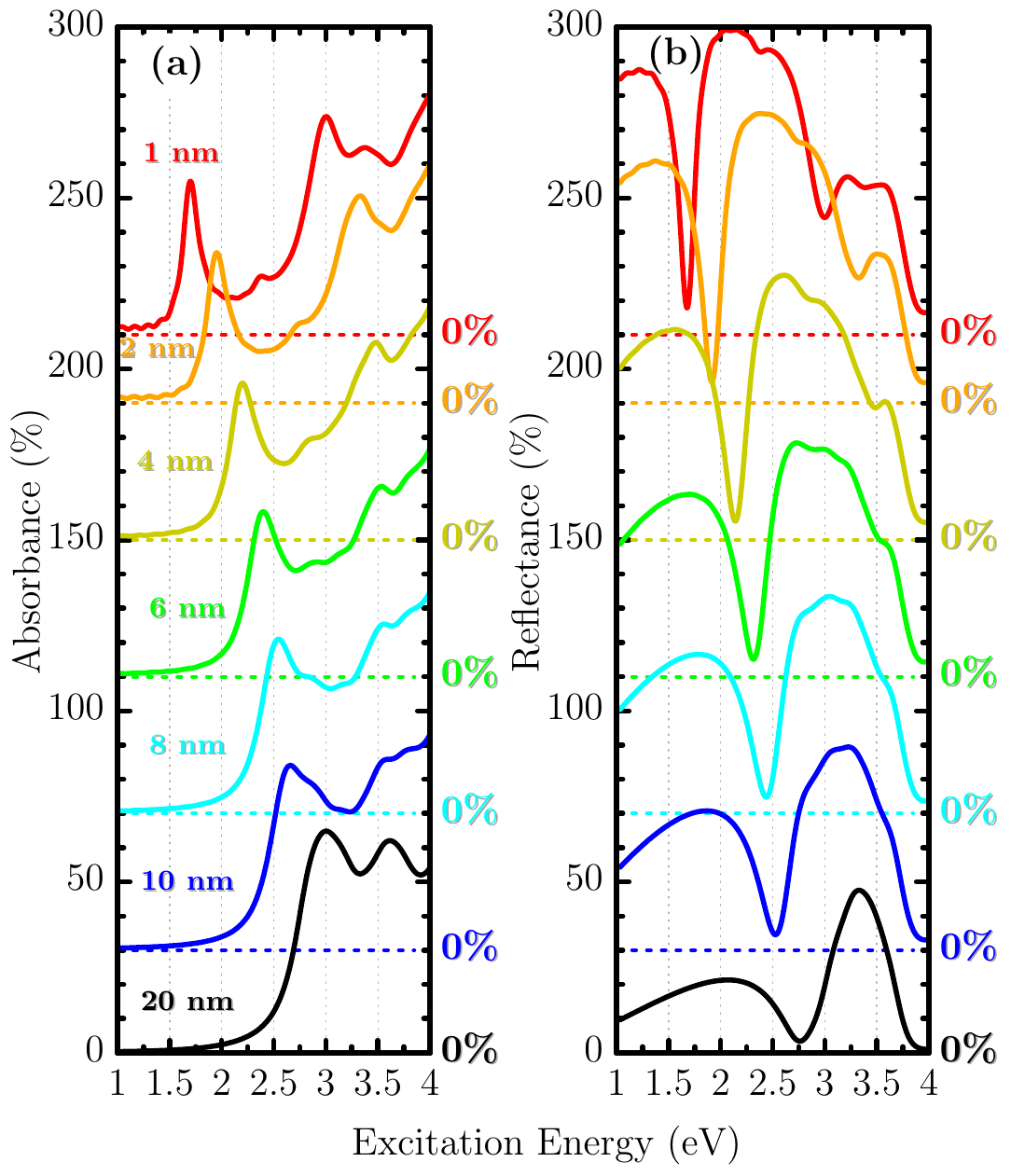}
	\caption{Simulated {\bfseries (a)} absorbance and {\bfseries (b)} reflectance spectra for silver nanoparticles separated by different gap sizes. The diameters were fixed to 50~nm. The color dashed lines indicate the offset of each spectrum.}
	\label{fig:Ag_NL2_SwAllGap_Waterfall}
\end{figure}

